\documentclass{article}
\usepackage[latin9]{inputenc}
\usepackage{geometry}
\usepackage{float}
\usepackage{amsmath}
\usepackage{amssymb}
\usepackage{graphicx}
\usepackage[numbers]{natbib}
\usepackage[unicode=true,
 bookmarks=false,
 breaklinks=false,pdfborder={0 0 1},backref=section,colorlinks=false]
 {hyperref}

\makeatletter
\usepackage{graphicx}          % Include figure files
\usepackage{amsmath}    % need for subequations
\usepackage{amssymb}    % for symbols
\usepackage{hyperref}   % use for hypertext links, including those to external documents and URLs
\usepackage[capitalise]{cleveref}   % use for referencing figures/equations

\usepackage{enumitem}
\usepackage{amsthm}
\usepackage{float}
\usepackage[capitalise]{cleveref}% use for referencing figures/equations
\usepackage{appendix}
\usepackage{color}
\usepackage{soul}
\usepackage{bm}

\makeatother

\begin{document}
\title{Machine learning for molecular simulation}
\author{Frank Noé$^{1,2,3,4,*}$, Alexandre Tkatchenko$^{6,\dagger}$, Klaus-Robert
Müller$^{7,8,9,\ddagger}$, Cecilia Clementi$^{1,3,4,5,\diamond}$}

\maketitle
$^{1}$ Freie Universität Berlin, Department of Mathematics and Computer
Science, Arnimallee 6, 14195 Berlin, Germany

$^{2}$ Freie Universität Berlin, Department of Physics, Arnimallee
6, 14195 Berlin, Germany

$^{3}$ Rice University, Department of Chemistry, Houston TX, 77005,
United States

$^{4}$ Rice University, Center for Theoretical Biological Physics,
Houston, Texas 77005, United States

$^{5}$ Rice University, Department of Physics, Houston TX, 77005,
United States

$^{6}$ Physics and Materials Science Research Unit, University of
Luxembourg, L-1511 Luxembourg, Luxembourg

$^{7}$ Technical University Berlin, Department of Computer Science,
Marchstr. 23, 10587 Berlin, Germany

$^{8}$ Max-Planck-Institut für Informatik, Saarbrücken, Germany

$^{9}$ Dept. of Brain and Cognitive Engineering, Korea University,
Seoul, South Korea

$*$: frank.noe@fu-berlin.de

$\dagger$: alexandre.tkatchenko@uni.lu

$\ddagger$: klaus-robert.mueller@tu-berlin.de

$\diamond$: cecilia@rice.edu

\vspace{0.5cm}

\begin{abstract}
Machine learning (ML) is transforming all areas of science. The complex
and time-consuming calculations in molecular simulations are particularly
suitable for a machine learning revolution and have already been profoundly
impacted by the application of existing ML methods. Here we review
recent ML methods for molecular simulation, with particular focus
on (deep) neural networks for the prediction of quantum-mechanical
energies and forces, coarse-grained molecular dynamics, the extraction
of free energy surfaces and kinetics and generative network approaches
to sample molecular equilibrium structures and compute thermodynamics.
To explain these methods and illustrate open methodological problems,
we review some important principles of molecular physics and describe
how they can be incorporated into machine learning structures. Finally,
we identify and describe a list of open challenges for the interface
between ML and molecular simulation.
\end{abstract}

\clearpage

\section{Introduction}

In 1929 Paul Dirac stated that:
\begin{quotation}
``The underlying physical laws necessary for the mathematical theory
of a large part of physics and the whole of chemistry are thus completely
known, and the difficulty is only that the exact application of these
laws leads to equations much too complicated to be soluble. It therefore
becomes desirable that approximate practical methods of applying quantum
mechanics should be developed, which can lead to an explanation of
the main features of complex atomic systems without too much computation.''
\citep{Dirac_1929}
\end{quotation}
Ninety years later, this quote is still state of the art. However,
in the last decade, new tools from the rapidly developing field of
machine learning (ML) have started to make significant impact on the
development of approximate methods for complex atomic systems, bypassing
the direct solution of \textquotedblleft equations much too complicated
to be soluble\textquotedblright .

ML aims at extracting complex patterns and relationships from large
data sets, to predict specific properties of the data. A classical
application of machine learning is to the problem of image classification
where descriptive labels need to be associated to images that are
presented in terms of sets of pixels. The \textquotedblleft machine\textquotedblright{}
is trained on a large number of examples and \textquotedblleft learns\textquotedblright{}
how to classify new images. The underlying idea is that there exists
a complex relationship between the input (the pixels) and the output
(the labels) that is unknown in its explicit form but can be inferred
by a suitable algorithm. Clearly, such an operating principle can
be very useful in the description of atomic and molecular systems
as well. We know that atomistic configurations dictate the chemical
properties, and the machine can learn to associate the latter to the
former without solving first principle equations, if presented with
enough examples.

Although different machine learning tools are available and have been
applied to molecular simulation (e.g., kernel methods \citep{Muelleretal2001}),
here we mostly focus on the use of neural networks, now often synonymously
used with the term ``deep learning''. We assume the reader has basic
knowledge of machine learning and we refer to the literature for an
introduction to statistical learning theory \citep{Vapnik_IEEE99_StatisticalLearningTheory,Bishop_Springer06_PatterRecognitionMachineLearningBook}
and deep learning \citep{LeCunBengioHinton_DeepLearning_Nature05,GoodfellowBengioCourville_BookDL}.

One of the first applications of machine learning in Chemistry has
been to extract classical potential energy surfaces from quantum mechanical
(QM) calculations, in order to efficiently perform molecular dynamics
(MD) simulations that can incorporate quantum effects. The seminal
work of Behler and Parrinello in this direction \citep{BehlerParrinello_PRL07_NeuralNetwork}
has opened the way to a now rapidly advancing area of research \citep{RuppEtAl_PRL12_QML,Csanyi-PRL,ChmielaEtAl_NatComm18_TowardExact,SmithIsayevRoitberg_ChemSci17_ANI,SmithEtAl_ANI-CC,brockherde2017bypassing,DTNN,BereauEtAl_JCP18_NoncovalentML}.
In addition to atomistic force fields, it has been recently shown
that, in the same spirit, effective molecular models at resolution
coarser than atomistic can be designed by ML \citep{John2017,ZhangEtAl_JCP18_DeePCG,WangEtAl_ACSCS19_CGnet}.
Analysis and simulation of MD trajectories has also been affected
by ML, for instance for the definition of optimal reaction coordinates
\citep{WehmeyerNoe_TAE,Hernandez_PRE18_VTE,MardtEtAl_VAMPnets,RibeiroTiwary_JCP18_RAVE,CheeSidkyFerguson_arxiv19_ReversibleVAMPnets,Jung2019},
the estimate of free energy surfaces \citep{Stecher_2014,Mones_2016,Schneider_2017,RibeiroTiwary_JCP18_RAVE},
the construction of Markov State Models \citep{MardtEtAl_VAMPnets,CheeSidkyFerguson_arxiv19_ReversibleVAMPnets,WuMardtPasqualiNoe_DeepGenerativeMSMs},
and for enhancing MD sampling by learning bias potentials \citep{ValssonParrinello_PRL14_VariationalMeta,BonatiZhangParrinello_arxiv19_NeuralVES,ZhangEtAl_Chemrxiv19_TALOS,McCartyParrinello_VACMetadynamics,SultanPande_JCTC17_TICAMetadynamics}
or selecting starting configurations by active learning \citep{DoerrDeFabritiis_JCTC14_OnTheFly,ZimmermanBowman_JCTC15_FAST,PlattnerEtAl_NatChem17_BarBar}.
Finally, ML can be used to generate samples from the equilibrium distribution
of a molecular system without performing MD altogether, as proposed
in the recently introduced Boltzmann Generators \citep{NoeEtAl_19_BoltzmannGenerators}.
A selection of these topics will be reviewed and discussed in the
following.

All these different aspects of molecular simulation have evolved independently
so far. For instance, ML-generated force-fields have mostly been developed
and applied on small molecules and ordered solids, while the analysis
of MD trajectories is mostly relevant for the simulation of large
flexible molecules, like proteins. In order to really revolutionize
the field, these tools and methods need to evolve, become more scalable
and transferable, and converge into a complete pipeline for the simulation
and analysis of molecular systems. There are still some significant
challenges towards this goal, as we will discuss in the following,
but considering the rapid progress of the last few years, we envision
that in the near future machine learning will have transformed the
way molecular systems are studied in silico.

This review focuses on physics-based ML approaches for molecular simulation.
ML is also having a big impact in other areas of Chemistry without
involving a physics-based model, for example by directly attempting
to make predictions of physicochemical or pharmaceutical properties
\citep{JimenezEtAl_JCIM18_KDEEP,SkalicEtAl_Bioinf18_LigVoxel,WuEtAl_ChemSci18_MoleculeNet,FeinbergEtAl_ACSCS18_PotentialNet},
or to designing materials and molecules with certain desirable properties
using generative learning \citep{GomezBombarelli_ACSCentral_AutomaticDesignVAE,PopovaaIsayevTropsha_SciAdv18_RLDrugDesign,WinterEtAl_MultiobjectiveOptimization,WinterEtAl_Translation_Chemistry2019}.
We refer the interested reader to other recent reviews on the subject
\citep{Butler_2018,Sanchez_Lengeling_2018}.

The review is organized as follows. We start by describing the most
important machine learning problems and principles for molecular simulation
(Sec. \ref{sec:ML_problems}). A crucial aspect of the application
of ML in molecular simulations is to incorporate physical constraints
and we will discuss this for the most commonly used physical symmetries
and invariances for molecular systems (Sec. \ref{sec:Physics}). We
will then provide examples of specific ML methods and applications
for molecular simulation tasks, focusing on deep learning and the
neural network architectures involved (Sec. \ref{sec:Architectures}).
We conclude by outlining open challenges and pointing out possible
approaches to their solution (Sec. \ref{sec:Discussion}).

\section{Machine Learning Problems for Molecular Simulation}

\label{sec:ML_problems}

In this section we discuss how several of the open challenges in the
simulation of molecular systems can be formulated as machine learning
problems, and the recent efforts to address them.

\subsection{Potential energy surfaces}

\label{subsec:MLProblem_PES}

Molecular dynamics (MD) and Markov Chain Monte Carlo (MCMC) simulations
employing classical force fields within the Born-Oppenheimer approximation
constitute the cornerstone of contemporary atomistic modeling in chemistry,
biology, and materials science. These methods perform importance sampling,
i.e. in the long run, they sample states $\mathbf{x}$ from the molecular
system's equilibrium distribution which has the general form:
\begin{equation}
\mu(\mathbf{x})\propto\mathrm{e}^{-u(\mathbf{x})}.\label{eq:equilibrium}
\end{equation}
The reduced potential $u(\mathbf{x})$ contains terms depending on
the thermodynamic constraints (e.g. fixed temperature, pressure, etc.).
In the canonical ensemble (fixed number of particles, volume and temperature),
$u(\mathbf{x})=U(\mathbf{x})/k_{B}T$ where $k_{B}T$ is the thermal
energy at temperature $T$.

However, the predictive power of these simulations is only as good
as the underlying potential energy surface (PES). Hence, predictive
simulations of properties and functions of molecular systems require
an accurate description of the global PES, $U_{{\rm }}({\bf x)}$,
where ${\bf {x}}$ indicates the nuclear Cartesian coordinates. All
many-body interactions between electrons are encoded in the $U(\mathbf{{x}})_{{\rm }}$
function. Although $U(\mathbf{x})_{{\rm }}$ could be obtained on
the fly using explicit ab initio calculations, more efficient approaches
that can access long time scales are required to understand relevant
phenomena in large molecular systems. A plethora of classical mechanistic
approximations to $U(\mathbf{x})_{{\rm }}$ exist, in which the parameters
are typically fitted to a small set of ab initio calculations or experimental
data \citep{LindorffLarsenEtAl_PlosOne12_SystematicForceField,Nerenberg_2012,Huang_2016,Robustelli_2018}.
Unfortunately, these classical approximations often suffer from the
lack of transferability and can yield accurate results only close
to the conditions (geometries) they have been fitted to.

\begin{figure}
\centering{}\includegraphics[width=0.8\textwidth]{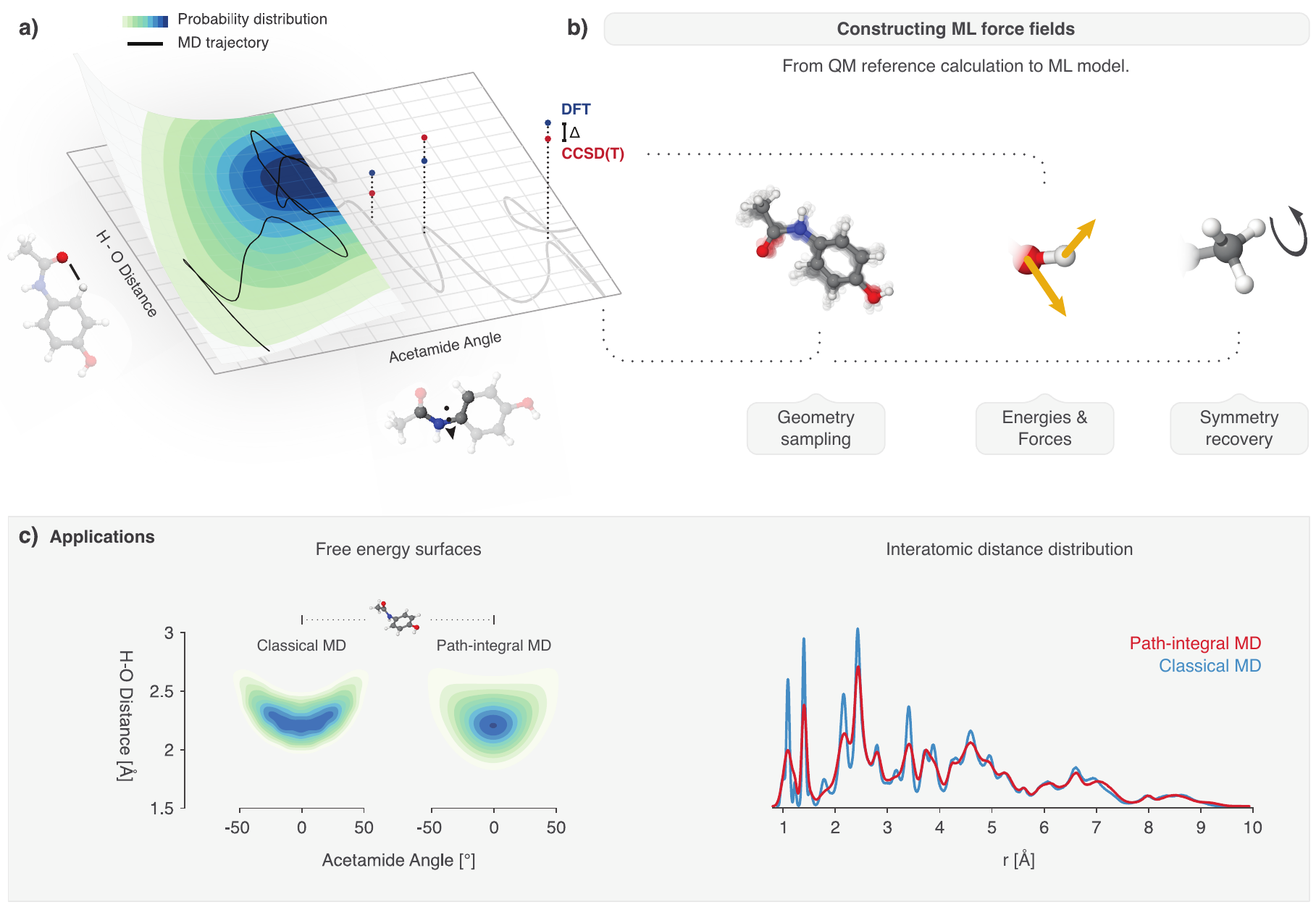}\caption{\label{fig:Constructing-force-field}Constructing force field models
with ML: \textbf{a}) Reference geometries are sampled from a MD trajectory
that is sufficiently long to ensure an optimal coverage of the configuration
space. \textbf{b}) For a small subset of geometries, energy and force
labels are then computed at a high level of theory to form the training,
validation and test datasets. A globally consistent atom-atom assignment
across the whole training set enables the identification and reconstructive
exploitation of relevant spatial and temporal physical symmetries
of the MD. \textbf{c}) The resulting ML model of the PES is finally
used to speed up sampling intensive path-integral MD simulations at
the accuracy of the reference electronic structure method \citep{ChmielaEtAl_NatComm18_TowardExact}.}
\end{figure}

Alternatively, sophisticated ML approaches that can accurately reproduce
the global potential energy surface (PES) for elemental materials
\citep{BehlerParrinello_PRL07_NeuralNetwork,Behler-review-JCP,Csanyi-PRL,Li-DeVita-PRL,DTNN,SchNet-JCP}
and small molecules \citep{Gastegger-ChemSci,Dral-JCP,SmithIsayevRoitberg_ChemSci17_ANI,ChmielaEtAl_NatComm18_TowardExact,ChmielaEtAl_SciAdv17_EnergyConserving,SchNet-JCP,HanZhangCarE_PRL18_DeepPotNet}
 have been recently developed (see Fig. \ref{fig:Constructing-force-field}).
Such methods learn a model of the PES, $\hat{U}(\mathbf{x},\theta)$,
where the parameters $\theta$ are optimized either by energy matching
and/or force matching. In energy matching, an ML model is trained
to minimize the loss function:

\begin{equation}
L_{\mathrm{ene}}=\sum_{i}(\hat{U}(\mathbf{x}_{i},\theta)-U_{i})^{2}\label{eq:loss_energy_matching}
\end{equation}
where $U_{i}$ are energy values obtained by QM calculations at specific
configurations (see Fig. \ref{fig:Constructing-force-field}). For
force matching, we compute the QM forces at specified configurations
\begin{equation}
\mathbf{f}(\mathbf{x})=\nabla U(\mathbf{x})\label{eq:conservative_force}
\end{equation}
and minimize the loss function:

\begin{equation}
L_{\mathrm{force}}=\sum_{i}\left\Vert \nabla\hat{U}(\mathbf{x}_{i},\theta)+\mathbf{f}_{i}\right\Vert ^{2}.\label{eq:loss_force_matching-1}
\end{equation}
The existing ML PES models are based either on non-linear kernel learning
\citep{Csanyi-PRL,Li-DeVita-PRL,ChmielaEtAl_SciAdv17_EnergyConserving,ChmielaEtAl_NatComm18_TowardExact}
or (deep) neural networks \citep{Behler-review-JCP,DTNN,HanZhangCarE_PRL18_DeepPotNet}.
Specific neural network architectures will be discussed in Sec. \ref{subsec:DeepLearningMD}
and \ref{subsec:SchNet}. Both approaches have advantages and limitations.
The advantage of the kernel methods is their convex nature yielding
a unique solution, whose behavior can be controlled outside of the
training data. Neural networks are non-convex ML models and often
harder to interpret and generalize outside of the training manifold.

\subsection{Free energy surfaces}

\label{subsec:MLProblem_FES}Given the Cartesian coordinates of a
molecule with $N$ atoms, $\mathbf{x\in}\mathbb{R}^{3N}$, we define
the collective coordinates by the encoding:
\begin{equation}
\mathbf{y}=E(\mathbf{x})\label{eq:CGmap}
\end{equation}
where $\mathbf{y}\in\mathbb{R}^{m}$ and $m$ is a small number. In
machine learning terms, the coordinates $\mathbf{y}$ define a \textit{latent
space.} In molecular simulations, the collective coordinates are often
defined to describe the slowest processes of the system \citep{NoeClementi_COSB17_SlowCVs}.
In general, the mapping (or encoding) $E$ can be highly non-linear.
If the energy function $U(\mathbf{x)}$ associated with the atomistic
representation is known, an important quantity to compute, for instance
to connect with experimental measurements, is the free energy of the
system as a function of the collective variables. The free energy
is defined as: 
\begin{equation}
F(\mathbf{y})=-\log\mu_{Y}(\mathbf{y})+\mathrm{const}\label{eq:free_energy}
\end{equation}
where $\mu_{Y}(\mathbf{y})$ is the marginal distribution on $\mathbf{y}$
of the equilibrium distribution $\mu(\mathbf{x})$ given by Eq. (\ref{eq:equilibrium}):
\begin{equation}
\mu_{Y}(\mathbf{y})=\int_{\mathbf{x}|M(\mathbf{x})=\mathbf{y}}\mu(\mathbf{x})\mathrm{d}\mathbf{x}.\label{eq:marginal_distribution}
\end{equation}
The integral in Eq. (\ref{eq:marginal_distribution}) is in practice
impossible to compute analytically for high dimensional systems, and
several methods have been developed for its numerical estimation by
enhancing the sampling of the equilibrium distribution in molecular
dynamics simulations of the system.

The definition of the free energy can also be formulated as a learning
problem: the aim is to optimize the parameters $\theta$ of a free
energy function model $\hat{F}(\mathbf{y},\theta)$ such that Eqs.
(\ref{eq:free_energy}-\ref{eq:marginal_distribution}) are satisfied
to a good approximation. Fulfilling these equations is usually referred
to as enforcing thermodynamic consistency.

Using methods that estimate the free energy $F^{\lambda}$ (or its
gradient $\nabla F^{\lambda}$) at a given set of points in the collective
variables space $\mathbf{y^{\lambda}}$ ($\lambda=1,\dots,M)$, one
can use the free energy loss:
\[
L_{ene}=\sum_{\lambda=1}^{M}\|F^{\lambda}-\hat{F}(\mathbf{y^{\lambda},\theta)}\|^{2}
\]
or the free energy gradient loss:
\[
L_{grad}=\sum_{\lambda=1}^{M}\|\nabla F^{\lambda}-\nabla\hat{F}(\mathbf{y^{\lambda},\theta)}\|^{2}
\]
with these free energy estimates in order to reconstruct the entire
surface $\hat{F}(\mathbf{y,}\theta)$. Both kernel regression \citep{Stecher_2014}
and neural networks \citep{Schneider_2017} have been used to this
purpose. Machine learning has also been used in combination with enhanced
sampling methods to learn the free energy surface on-the-fly \citep{Mones_2016,Galvelis_2017,Sidky_2018,RibeiroTiwary_JCP18_RAVE,Lamim_Ribeiro_2018,Chen_2018,Sultan_2018,Guo_2018}.

An alternative way to learn (\ref{eq:free_energy}) is by using force
matching \citep{Noid2008,Noid2013}. It has been demonstrated \citep{Ciccotti_2008}
that, given the forces $\nabla_{\mathbf{x}}U$ of the atomistic system
collected along a molecular dynamic trajectory, $\mathbf{x}_{t}$,
$t=1,\dots,T$, the gradient of a free energy model $\hat{F}(\mathbf{y,}\theta)$
that best satisfies Eq. (\ref{eq:free_energy}) also minimizes the
force matching loss:

\begin{align}
L_{\mathrm{force}}=\sum_{t}\left\Vert \nabla_{\mathbf{y}}\hat{F}(E(\mathbf{x}_{_{t}}),\theta)+f_{lmf}^{\mathbf{y}}(\mathbf{x}_{t})\right\Vert ^{2}\label{eq:loss_force_matching2}
\end{align}
where the term $f_{lmf}^{\mathbf{y}}$ is the \textit{local mean force:
\begin{align*}
f_{lmf}^{\mathbf{y}}(\mathbf{x}) & =\nabla_{\mathbf{x}}U\cdot G^{\mathbf{y}}+\nabla_{\mathbf{x}}\cdot G^{\mathbf{y}}\\
G^{\mathbf{y}} & =\nabla_{\mathbf{x}}E[(\nabla_{\mathbf{x}}E)^{T}\nabla_{\mathbf{x}}E]^{-1}
\end{align*}
}that is, the projection of the atomistic force $\nabla_{\mathbf{x}}U$
on the collective variable space through the mapping $E$.

In practice, the estimator (\ref{eq:loss_force_matching2}) is very
noisy: because of the dimensionality reduction from $\mathbf{x}$
to $\mathbf{y}$, multiple realizations of the projected force $f_{lmf}^{\mathbf{y}}$
can be associated to the same value of the collective coordinates
$\mathbf{y}$ and the minimum of the loss function (\ref{eq:loss_force_matching2})
can not go to zero. By invoking statistical estimator theory it can
be shown that this loss can be broken down into a bias, variance and
noise terms \citep{WangEtAl_ACSCS19_CGnet}.

\subsection{Coarse-graining}

\label{subsec:MLProblem_CG}

The use of coarse-grained models of complex molecular systems, such
as proteins, presents an attractive alternative to atomistic models
that are very expensive to simulate \citep{ClementiCOSB,Noid2013}.

The design of a coarse-grained model for a system with $N$ atoms
into a reduced representation with $n$ effective ``beads'' starts
by the definition of a mapping similar to Eq. (\ref{eq:CGmap}) where
now $\mathbf{y}\in\mathbb{R}^{3n}$ are the coordinates of the coarse-grained
beads. In this case, the mapping is usually a linear function, as
in general the beads are defined as a subset or a linear combination
of sets of atoms in the original system, in a way that allows to keep
some information about the geometry of the molecule. For instance,
in protein systems, a coarse-grained mapping can be defined by replacing
all the atoms in a residue by a bead centered on the corresponding
$C_{\alpha}$ atom. There is at present no general theory to define
the optimal coarse-graining mapping for a specific system. A few methods
have been proposed in this direction, by optimizing the mapping $E$
to preserve certain properties of the original system. Examples include
the definition of a system's dynamical assembly units \citep{BoninsegnaBanish2018}
or the use of an autoencoder to minimize the reconstruction error
\citep{WangBombarelli_Autograin}.

Once the coarse-graining mapping is given, several strategies exist
to define the model energy function, either to match experimental
observables (top-down), or to reproduce specific properties of the
atomistic system (bottom-up) \citep{Noid2013}. If one wants to define
a coarse-grained model that is thermodynamically consistent with the
original model (Eqs. \ref{eq:free_energy}-\ref{eq:marginal_distribution}),
then, the definition of the energy function associated with a coarse-grained
molecular model can be seen as a special case of free energy learning
discussed in the previous section \citep{Noid2008}. The effective
energy of the coarse-grained model is the free energy defined by Eq.
(\ref{eq:free_energy}) where $\mathbf{y}$ are now the coarse variables.
Therefore, once the coarse-graining map is defined, the definition
of the effective potential can also be seen as a learning problem.
In particular, the force matching loss function given by Eq. (\ref{eq:loss_force_matching2})
can be used to train the effective energy of the model from the atomistic
forces. In the case of a linear mapping with a crisp assignment of
atoms into beads, a matrix $\mathbf{M}\in\mathbb{R}^{3n\times3N}$
exists such that $\mathbf{y}=\mathbf{M}\mathbf{x}$ and the expression
for the loss, Eq. (\ref{eq:loss_force_matching2}), reduces to:
\begin{align}
L_{force} & =\sum_{t}\|\nabla_{\mathbf{y}}\hat{F}(\mathbf{M}\mathbf{x}_{t},\theta)+f(\mathbf{x}_{t})\|^{2}\label{eq:loss-force-matching_linear}
\end{align}
where for each bead $I$, $f_{I}=\sum_{i\in I}\nabla_{\mathbf{x}_{i}}U$
is the sum of the atomistic forces $\nabla_{\mathbf{x}_{i}}U$ of
all the atoms mapping to that bead. This loss function has been used
to design coarse grained force fields for different systems both with
kernel methods \citep{John2017} and deep neural networks \citep{ZhangEtAl_JCP18_DeePCG,WangEtAl_ACSCS19_CGnet}.

\subsection{Kinetics}

\label{subsec:MLProblem_Kinetics}

Kinetics are the slow part of the dynamics. Due to the stochastic
components in the MD integrator, for any trajectory emerging from
a configuration $\mathbf{x}_{t}$ at time $t$, there is a probability
distribution of finding the molecule in configuration $\mathbf{x}_{t+\tau}$
at a later time $t+\tau$: 
\begin{equation}
\mathbf{x}_{t+\tau}\sim p_{\tau}(\mathbf{x}_{t+\tau}\mid\mathbf{x}_{t}).\label{eq:transition_density}
\end{equation}
We can express the transition density (\ref{eq:transition_density})
as the action of the Markov propagator in continuous-space, and by
its spectral decomposition \citep{SarichNoeSchuette_MMS09_MSMerror,WuNoe_VAMP}:
\begin{align}
p(\mathbf{x}_{t+\tau}) & =\int p(\mathbf{x}_{t+\tau}\mid\mathbf{x}_{t};\tau)p(\mathbf{x}_{t})\:\mathrm{d}\mathbf{x}_{t}\approx\sum_{k=1}^{n}\sigma_{k}^{*}\langle p(\mathbf{x}_{t})\mid\phi(\mathbf{x}_{t})\rangle\psi(\mathbf{x}_{t+\tau}).\label{eq:density_propagation}
\end{align}
The spectral decomposition can be read as follows: The evolution of
the probability density $p(\mathbf{x})$ can be approximated as the
superposition of functions $\psi$. A second set of functions, $\phi$,
is required in order to compute the amplitudes of these functions.

In general, Eq. (\ref{eq:density_propagation}) is a singular value
decomposition with left and right singular functions $\phi_{k},\psi_{k}$
and true singular values $\sigma_{k}^{*}$ \citep{WuNoe_VAMP}. The
approximation then is a low-rank decomposition in which the small
singular values are discarded. For the special case that dynamics
are in thermal equilibrium, Eq. (\ref{eq:detailed_balance}) holds,
and Eq. (\ref{eq:density_propagation}) is an eigenvalue decomposition
with the choices:
\begin{align}
\sigma_{k}^{*} & =\lambda_{k}^{*}(\tau)=\mathrm{e}^{-\tau/t_{i}^{*}}\label{eq:timescale}\\
\phi_{k}(\mathbf{x}) & =\psi_{k}(\mathbf{x})\mu(\mathbf{x}).\nonumber 
\end{align}
Hence Eq. (\ref{eq:density_propagation}) simplifies: we only need
one set of functions, the eigenfunctions $\psi_{k}$. The true eigenvalues
$\lambda_{k}^{*}$ are real-valued and decay exponentially in time
$\tau$ with the characteristic relaxation times $t_{i}^{*}$ that
are directly linked to kinetic experimental observables \citep{BucheteHummer_JPCB08,NoeEtAl_PNAS11_Fingerprints}.
The approximation in Eq. (\ref{eq:density_propagation}) is due to
truncating all terms with relaxation times shorter than $t_{n}^{*}$.

A quite established approach is to learn molecular kinetics (Eq. \ref{eq:density_propagation})
from a given trajectory dataset. In order to obtain a low-dimensional
model of the molecular kinetics that is easy to interpret and analyze,
this usually involves two steps: (i) finding a low-dimensional latent
space representation of the collective variables, $\mathbf{y}=E(\mathbf{x})$,
using the encoder $E$, and (ii) learning a dynamical propagator $\mathbf{P}$
in that space:
\begin{equation}
\begin{array}{ccc}
\text{\ensuremath{\mathbf{x}}}_{t} & \overset{E}{\longrightarrow} & \text{\ensuremath{\mathbf{y}}}_{t}\\
\mathrm{MD}\downarrow\:\:\:\:\:\:\:\:\:\:\: &  & \:\:\:\:\:\downarrow\mathbf{P}\\
\text{\ensuremath{\mathbf{x}}}_{t+\tau} & \overset{E}{\longrightarrow} & \text{\ensuremath{\mathbf{y}}}_{t+\tau}
\end{array}\label{eq:learningDynamics}
\end{equation}
A common approach in MD, but also in other fields such as dynamical
systems and fluid mechanics, is to seek a pair $(E,\mathbf{P})$,
such that $\mathbf{P}$ is a small matrix that propagates state vectors
in a Markovian (memoryless) fashion \citep{SchuetteFischerHuisingaDeuflhard_JCompPhys151_146,SwopePiteraSuits_JPCB108_6571,NoeHorenkeSchutteSmith_JCP07_Metastability,ChoderaEtAl_JCP07,BucheteHummer_JPCB08,PrinzEtAl_JCP10_MSM1,Mezic_NonlinDyn05_Koopman,SchmidSesterhenn_APS08_DMD,TuEtAl_JCD14_ExactDMD,WuEtAl_JCP17_VariationalKoopman,WuNoe_VAMP}.
This is motivated by the spectral decomposition of dynamics (Eq. \ref{eq:density_propagation}):
If $\tau$ is large enough to filter fast processes, a few functions
are sufficient to describe the kinetics of the system, and if $E$
maps to the space spanned by these functions, $\mathbf{P}$ can be
a linear, Markovian model.

If no specific constraints are imposed on $\mathbf{P}$, the minimum
regression error of $\mathbf{y}_{t+\tau}=\mathbf{P}\mathbf{y}_{t}$,
the variational approach of Markov processes (VAMP) \citep{NoeNueske_MMS13_VariationalApproach,WuNoe_VAMP},
or maximum likelihood will all lead to the same estimator \citep{Noe_LectureNotes_ML4MD}:
\begin{equation}
\mathbf{P}=\mathbf{C}_{00}^{-1}\mathbf{C}_{0\tau},\label{eq:transition_matrix_estimator-1}
\end{equation}
using the latent space covariance matrices
\[
\begin{array}{ccc}
\mathbf{C}_{00}=\frac{1}{T}\sum_{t=1}^{T-\tau}\mathbf{y}_{t}\mathbf{y}_{t}^{\top} & \mathbf{C}_{0\tau}=\frac{1}{T}\sum_{t=1}^{T-\tau}\mathbf{y}_{t}\mathbf{y}_{t+\tau}^{\top} & \mathbf{C}_{\tau\tau}=\frac{1}{T}\sum_{t=1}^{T-\tau}\mathbf{y}_{t+\tau}\mathbf{y}_{t+\tau}^{\top}\end{array}.
\]

If $E$ performs a one-hot-encoding that indicates which ``state''
the system is in, then the pair $(E,\mathbf{P})$ is called Markov
state model (MSM \citep{SchuetteFischerHuisingaDeuflhard_JCompPhys151_146,SwopePiteraSuits_JPCB108_6571,NoeHorenkeSchutteSmith_JCP07_Metastability,ChoderaEtAl_JCP07,BucheteHummer_JPCB08,PrinzEtAl_JCP10_MSM1}),
and $\mathbf{P}$ is a matrix of conditional probabilities to be in
a state $j$ at time $t+\tau$ given that the system was in a state
$i$ at time $t$.

Learning the embedding $E$ is more difficult than $\mathbf{P}$,
as optimizing $E$ by maximum likelihood or minimal regression error
in latent space $\mathbf{y}$ leads to a collapse of $E$ to trivial,
uninteresting solutions \citep{Noe_LectureNotes_ML4MD}. This problem
can be avoided by following a variational approach to approximating
the leading terms of the spectral decomposition (\ref{eq:transition_density})
\citep{NoeNueske_MMS13_VariationalApproach,WuNoe_VAMP}. The Variational
Approach for Conformation dynamics (VAC) \citep{NoeNueske_MMS13_VariationalApproach}
states that for dynamics obeying detailed balance (\ref{eq:detailed_balance}),
the eigenvalues $\lambda_{k}$ of a propagator matrix $\mathbf{P}$
via any encoding $\mathbf{y}=E(\mathbf{x})$ are, in the statistical
limit, lower bounds of the true $\lambda_{k}^{*}$. The VAMP variational
principle is more general, as it does not require detailed balance
(\ref{eq:detailed_balance}), and applies to the singular values $\sigma_{k}$:
\begin{align*}
\lambda{}_{k} & \le\lambda_{k}^{*}\:\:\:(\mathrm{with}\:\mathrm{detailed\:balance})\\
\sigma_{k} & \le\sigma_{k}^{*}\:\:\:(\mathrm{no}\:\mathrm{detailed\:balance}).
\end{align*}
As VAMP is the more general principle, we can use it to define the
loss function for estimating molecular kinetics models:

\begin{align}
\mathcal{L}_{\mathrm{VAMP}-2}(\{\mathbf{y}_{t}^{0},\mathbf{y}_{t}^{\tau}\}) & =-\left\Vert \left(\mathbf{C}_{00}\right)^{-\frac{1}{2}}\mathbf{C}_{0\tau}\left(\mathbf{C}_{\tau\tau}\right)^{-\frac{1}{2}}\right\Vert _{F}^{2}.\label{eq:VAMP2_loss_b}
\end{align}
If dynamics obey detailed balance, we can use $\mathbf{C}_{\tau\tau}=\mathbf{C}_{00}$
and plug in a symmetric estimate for $\mathbf{C}_{0\tau}$ \citep{Noe_LectureNotes_ML4MD,CheeSidkyFerguson_arxiv19_ReversibleVAMPnets}.

\subsection{Sampling and Thermodynamics}

\label{subsec:MLProblem_Thermodynamics}

MD time steps are on the order of one femtosecond ($10^{-15}$ s),
while configuration changes governing molecular function are often
rare events that may take $10^{-3}$ to $10^{3}$ s. Even when evaluating
the potential and the forces it generates is fast, simulating single
protein folding and unfolding events by direct MD simulation may take
years to centuries on a supercomputer. To avoid this sampling problem,
machine learning methods can be employed to learn generating equilibrium
samples from $\mu(\mathbf{x})$ more efficiently, or even by generating
statistically independent samples in ``one shot'' (Sec. \ref{subsec:MLProblem_Thermodynamics}).

Learning to sample probability distributions is the realm of generative
learning \citep{Smolensky_RBM,KingmaWelling_ICLR14_VAE,GoodfellowEtAl_GANs,DinhDruegerBengio_NICE2015}.
In the past few years, directed generative networks, such as variational
Autoencoders (VAEs) \citep{KingmaWelling_ICLR14_VAE}, generative
adversarial networks (GANs) \citep{GoodfellowEtAl_GANs}, and flows
\citep{TabakVandenEijnden_CMS10_DensityEstimation,DinhDruegerBengio_NICE2015},
have had a particular surge of interest. Such networks are trained
to transform samples from an easy-to-sample probability distribution,
such as Gaussian noise, into realistic samples of the objects of interest.
These methods have been used to draw photorealistic images \citep{KarrasEtAl_ProgressiveGrowingGANs,KingmaDhariwal_NIPS18_Glow},
generate speech or music audio \citep{VanDenOord_WaveNet2} and generate
chemical structures to design molecules or materials with certain
properties \citep{GomezBombarelli_ACSCentral_AutomaticDesignVAE,PopovaaIsayevTropsha_SciAdv18_RLDrugDesign,WinterEtAl_MultiobjectiveOptimization,WinterEtAl_Translation_Chemistry2019}.

If the aim is to learn a probability distribution from sampling data
(density estimation) or learn to efficiently sample from a given probability
distribution (Boltzmann generation \citep{NoeEtAl_19_BoltzmannGenerators}),
one typically faces the challenge of matching the probability distribution
of the trained model with a reference.

Matching probability distributions can be achieved by minimizing probability
distances. The most commonly used probability distance is the Kullback-Leibler
(KL) divergence, also called relative entropy between two distributions
$q$ and $p$:
\begin{align}
\mathrm{KL}(q\parallel p) & =\int q(\mathbf{x})\left[\log q(\mathbf{x})-\log p(\mathbf{x})\right]\mathrm{d}\mathbf{x}.\label{eq:KL}
\end{align}
In Boltzmann generation \citep{NoeEtAl_19_BoltzmannGenerators}, we
aim to efficiently sample the equilibrium distribution $\mu(\mathbf{x})$
of a many-body system defined by its energy function $u(\mathbf{x})$
(Eq. \ref{eq:equilibrium}). We can learn the parameters $\boldsymbol{\theta}$
of a neural network that generates samples from the distribution $p_{X}(\mathbf{x};\boldsymbol{\theta})$
by making its generated distribution similar to the target equilibrium
distribution. Choosing $q\equiv p_{X}(\mathbf{x})$ and $p\equiv\mu(\mathbf{x})$
in Eq. (\ref{eq:KL}), we can minimize the $\mathrm{KL}_{\boldsymbol{\theta}}\left[q_{X}\parallel\mu_{X}\right]$
with the energy loss:
\begin{equation}
\mathrm{EL}=\mathbb{E}_{\mathbf{x}\sim p_{X}(\mathbf{x};\boldsymbol{\theta})}\left[\log p_{X}(\mathbf{x};\boldsymbol{\theta})+u(\mathbf{x})\right].\label{eq:equilibrium_KLloss}
\end{equation}
As the network samples from an energy surface $u(\mathbf{x};\boldsymbol{\theta})=-\log p_{X}(\mathbf{x};\boldsymbol{\theta})+\mathrm{const}$
(Eq. \ref{eq:equilibrium}), this loss is performing energy matching,
similar as in Sec. \ref{subsec:MLProblem_PES}. In order to evaluate
the loss (\ref{eq:equilibrium_KLloss}), we need not only to be able
to generate samples $\mathbf{x}\sim p_{X}(\mathbf{x};\boldsymbol{\theta})$
from the network, but also to be able to compute $p_{X}(\mathbf{x};\boldsymbol{\theta})$
for every sample $\mathbf{x}$. See Sec. \ref{subsec:BGs} for an
example.

The reverse case is density estimation. Suppose we have simulation
data $\mathbf{x}$, and we want to train the probability distribution
$p_{X}(\mathbf{x};\boldsymbol{\theta})$ to resemble the data distribution.
We choose $q(\mathbf{x})$ as the data distribution and $p\equiv p_{X}(\mathbf{x};\boldsymbol{\theta})$
in Eq. (\ref{eq:KL}), and exploit that $\mathbb{E}_{\mathbf{x}\sim data}\left[\log q(\mathbf{x})\right]$
is a constant that does not depend on the parameters $\boldsymbol{\theta}$.
Then we can minimize the loss:
\begin{equation}
\mathrm{NL}=-\mathbb{E}_{\mathbf{x}\sim\mathrm{data}}\left[\log p_{X}(\mathbf{x};\boldsymbol{\theta})\right].\label{eq:equilibrium_MLloss}
\end{equation}
This loss is the negative likelihood that the model generates the
observed sample, hence minimizing it corresponds to a maximum likelihood
approach. Likelihood maximization is abundantly used in machine learning,
in this review we will discuss it in Sec. \ref{subsec:BGs}.

\section{Incorporating Physics into Machine Learning}

\label{sec:Physics}

\subsection{Why incorporate physics?}

Compared to classical ML problems such as image classification, we
have a decisive advantage when working with molecular problems: we
know a lot of physical principles that restrict the possible predictions
of our machine to the physically meaningful ones.

Let us start with a simple example to illustrate this. We are interested
in predicting the potential energy and the atom-wise forces of the
diatomic molecule $O_{2}$ with positions $\mathbf{x}_{1},\mathbf{x}_{2}$
in vacuum (without external forces). Independent of the details of
our learning algorithm, physics tells us that a few general rules
must hold:
\begin{enumerate}
\item The energy is invariant when translating or rotating the molecule.
We can thus arbitrarily choose the positions $\mathbf{x}_{1}=(0,0,0);\mathbf{x}_{2}=(d,0,0)$,
and the energy becomes a function of the interatomic distance only:
$U(\mathbf{x})\rightarrow U(d)$.
\item Energy conservation: the energy $U$ and the force of the molecule
are related by $\mathbf{F}(\mathbf{x})=-\nabla U(\mathbf{x})$. Now
we can compute the components of the force as: $\mathbf{f}_{1}=(\frac{\partial U(d)}{\partial d},0,0);\mathbf{f}_{2}=(-\frac{\partial U(d)}{\partial d},0,0)$.
\item Indistinguishability of identical particles: The energy is unchanged
if we exchange the labels ``1'' and ``2'' of the identical oxygen
atoms.
\end{enumerate}
In machine learning, there are two principal approaches when dealing
with such invariances or symmetries: (i) Data augmentation, and (ii)
Building the invariances into the ML model.

\subsection{Data augmentation}

Data augmentation means we learn invariances ``by heart'' by artificially
generating more training data and applying the known invariances to
it. For example, given a training data point for positions and energy/force
labels, $(\mathbf{x};U,\mathbf{f})$, we can augment this data point
with translation invariance of energy and force by adding more training
data $(\mathbf{x}+\Delta\mathbf{x};U,\mathbf{f})$ with random displacements
$\Delta\mathbf{x}$. Data augmentation makes the ML model more robust
and will help to predicting the invariances approximately. It is an
important ML tool, as it is easy to do, while for certain invariances
it is conceptually difficult or computationally expensive to hard-wire
them into the ML model.

However, data augmentation is statistically inefficient, as additional
training data are needed, and inaccurate because a network that does
not have translation invariance hard-wired into it will never predict
that the energy is \emph{exactly} constant when translating the molecule.
This inaccuracy may lead to unphysical, and potentially catastrophic
predictions when such energy model is inserted into an MD integrator.

\subsection{Building physical constraints into the ML model}

The more accurate, statistical efficient and also more elegant approach
to incorporating physical constraints is to directly build them into
the ML model. Doing so involves accounting for two related aspects
that are both essential to make the learning problem efficient: (i)
equivariances: the ML model should have the same invariances and symmetries
as the modeled physics problem, (ii) parameter sharing: whenever there
is an invariance or symmetry, this should be reflected by sharing
model parameters in different parts of the network.

\subsection{Invariance and Equivariance}

\label{subsec:Invariance-and-Equivariance}

\begin{figure}
\begin{centering}
\includegraphics[width=0.8\columnwidth]{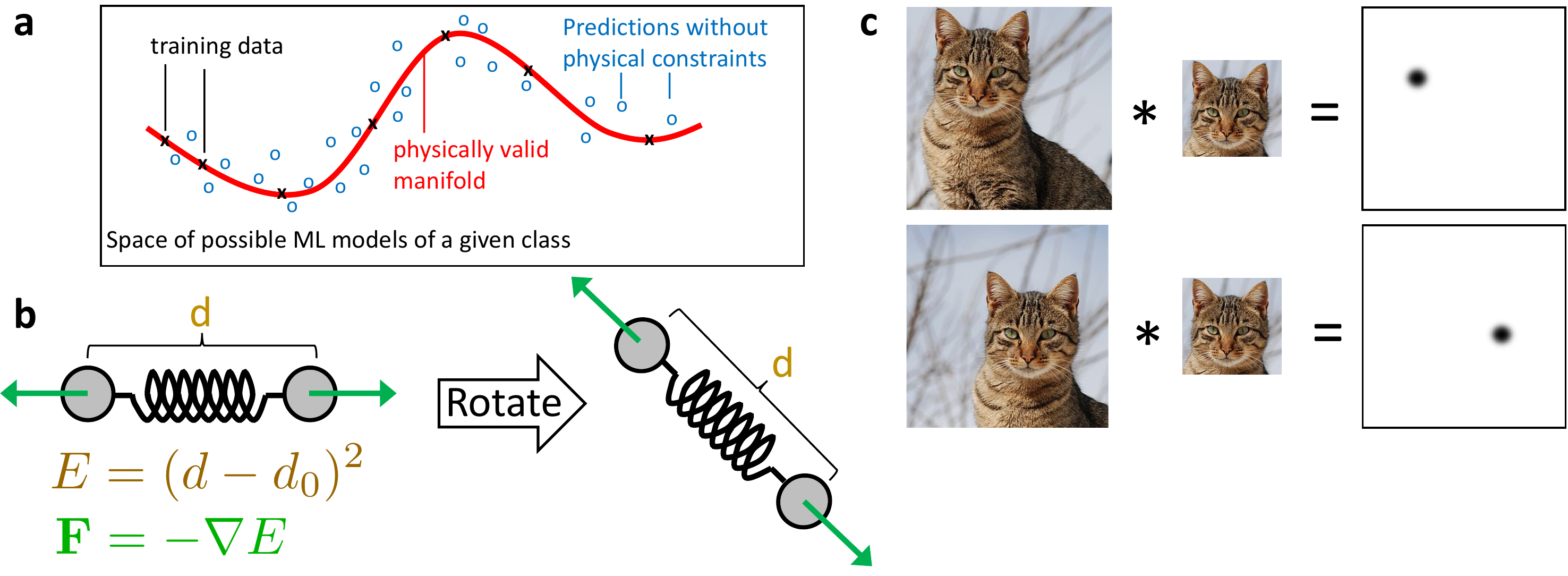}
\par\end{centering}
\caption{\label{fig:physical_constraints}\textbf{Physical constraints} \textbf{a})
Physical constraints define a manifold of physically valid solutions
for a given ML model class, e.g. a given neural network architecture:
only certain combinations of parameters will obey these physical constraints.
Using data augmentation, the network can learn ``by heart'' to make
almost physically valid predictions. Directly building physical constraints
into the ML model is more data-efficient and accurate, as then every
prediction is physically meaningful. \textbf{b})\textbf{ }Invariance
and equivariance: Upon rotating the beads, the spring energy is invariant,
but the forces of the beads are equivariant, i.e. they rotate in the
same way. \textbf{c}) Translational equivariance in convolutional
layers: convolving a translated image with a filter results in a translated
feature map. Modified from https://en.wikipedia.org/wiki/File:Cat\_November\_2010-1a.jpg.}
\end{figure}

If we can hard-wire the physical symmetries into the ML structure,
we can reduce the dimensionality of the problem. In the example of
the $O_{2}$ molecule above, the energy and force are one- and six-dimensional
functions defined for each point $\mathbf{x}$ in 6-dimensional configuration
space. However, when we use the invariances described above we only
have to learn a one-dimensional energy function over a one-dimensional
space, and can compute the full force field from it. The learning
problem has become much simpler because we now learn only in the manifold
of physically meaningful solutions (Fig. \ref{fig:physical_constraints}a).

An important concept related to invariance is equivariance. A function
is equivariant if it transforms the same way as its argument. For
example, the force is defined by the negative gradient of the energy,
$-\nabla U(\mathbf{x})$. If we rotate the molecule by applying the
rotation $\mathbf{R}\cdot$, the energy is invariant, but the force
is equivariant as it rotates in the same way as $\mathbf{x}$ (Fig.
\ref{fig:physical_constraints}b):
\[
\begin{array}{ccc}
\mathbf{x} & \overset{\mathbf{R\cdot}}{\longrightarrow} & \mathbf{R}\mathbf{x}\\
\:\:\:\:\downarrow\nabla U(\cdot) &  & \:\:\:\:\downarrow\nabla U(\cdot)\\
\nabla U(\mathbf{x}) & \overset{\mathbf{R}\cdot}{\longrightarrow} & \mathbf{R}\nabla U(\mathbf{x})=\nabla U(\mathbf{R}\mathbf{x})
\end{array}
\]
Equivariances are closely linked to convolutions in machine learning.
For example, standard convolutions are translation invariant: Each
convolution kernel is a feature detector that is applied to each pixel
neighborhood \citep{LeCun_ProcIEEE89_ConvNets}. When that feature
is present in the image, the convolved image will show a signal in
the corresponding position (Fig. \ref{fig:physical_constraints}c).
When translating the input, the convolved image translates in the
same way (Fig. \ref{fig:physical_constraints}c).

We will briefly review below common invariances/equivariances useful
for applications to molecular simulations, and discuss strategies
to incorporate them in an ML algorithm.

\subsubsection{Rototranslational invariance/equivariance}

\label{subsec:Rototranslational_invariance}

Physical quantities that only depend on the interactions of the atoms
within a molecule should be invariant with respect to translation
$\mathbf{T}$ and rotation $\mathbf{R}\cdot$. Examples include potential
and free energies of a molecule without an external field:
\[
U(\mathbf{R}\mathbf{x}+\mathbf{T})=U(\mathbf{x}).
\]
The force is equivariant to rotation, but invariant to translation
(Fig. \ref{fig:physical_constraints}b, Sec. \ref{subsec:Invariance-and-Equivariance}):
\begin{align*}
-\nabla U(\mathbf{R}\mathbf{x}+\mathbf{T}) & =-\mathbf{R}\nabla U(\mathbf{x})
\end{align*}
Rototranslational invariance can be achieved by transforming the configuration
$\mathbf{x}$ into roto-translationally invariant features, such as
intramolecular distances or angles. Equivariance of the force can
then be achieved by computing the force explicitly by a network layer
that computes the derivatives with respect to $\mathbf{x}$, as it
is done, e.g., in SchNet \citep{Nipsschnet} and CGnet \citep{WangEtAl_ACSCS19_CGnet}
(Fig. \ref{fig:CGnet}a). Note that periodic systems, such as crystals
and explicit-solvent boxes have translational but not rotational invariances/equivariances.

\subsubsection{Permutational invariance/equivariance.}

\label{subsec:Permutational_invariance}

Physical quantities, such as quantum mechanical energies, are invariant
if we exchange the labels of identical atoms, e.g., Carbons. As the
number of possible permutations increases exponentially with the number
of identical particles, trying to learn permutation invariance by
data augmentation is hopeless. Permutation invariance can be built
into the ML model by choosing a functional form to compute the quantity
of interest that is permutation invariant -- see \citep{ZaheerEtAl_NIPS2017_DeepSets}
for the general conditions. Following pioneering work of \citep{Ercolessi_PRL86_Au100,Tersoff_TRL86_Silicon,FerranteEtAl_PRL83_DiatomicMetallic,Abell_PRB85_Pseudopotential},
a specific choice that is common for networks that compute extensive
quantities such as potential energies $U(\mathbf{x})$ is to model
it as a sum
\begin{equation}
U(\mathbf{x})=\sum_{i}U_{i}(\mathbf{x}),\label{eq:energy_atomwise}
\end{equation}
where $U_{i}$ is the contribution of the energy by the $i$th atom
in its chemical environment \citep{BehlerParrinello_PRL07_NeuralNetwork,SmithIsayevRoitberg_ChemSci17_ANI,HanZhangCarE_PRL18_DeepPotNet,Nipsschnet}.
In order to account for the multi-body character of quantum mechanics,
$U_{i}(\mathbf{x})$ must generally also be a multi-body function.
Eq. (\ref{eq:energy_atomwise}) can be implemented by using separate
networks for the individual contributions $U_{i}$ and adding up their
results (Fig. \ref{fig:Behler-Parrinello}). The force resulting from
Eq. (\ref{eq:energy_atomwise}) is automatically permutation equivariant.

Classical MD force fields define bonded interactions by assigning
atoms to nodes in a bond graph, and thus exchanging of individual
atom pairs no longer preserves the energy. However, the energy is
still invariant to the exchange of identical molecules, e.g., solvent,
and is important to take that into account for coarse-graining \citep{WangEtAl_ACSCS19_CGnet}
and generating samples from the equilibrium density \citep{NoeEtAl_19_BoltzmannGenerators}.

A simple alternative to building permutation invariance into the ML
function is to map all training and test data to a reference permutation.
This can be done efficiently by so-called bipartite graph matching
methods such as the Hungarian method \citep{Kuhn_Naval55_HungarianMethod},
that are frequently used in recent learning models \citep{Reinhard_2007,ChmielaEtAl_NatComm18_TowardExact,NoeEtAl_19_BoltzmannGenerators}.

\subsubsection{Energy conservation}

Closed physical systems are energy-conserving, which implies that
the force field is defined by the gradient of the energy (Eq. \ref{eq:conservative_force}).
When learning potential energy, it can be a great advantage to use
force information, because each $N$-atom configuration is associated
with only one energy but $3N$ forces, which may result in superior
data efficiency when force labels are used during learning \citep{ChmielaEtAl_SciAdv17_EnergyConserving,ChmielaEtAl_NatComm18_TowardExact}.
If we use supervised learning for coarse-graining with thermodynamic
consistency, we can only use forces, as no labels for the free energies
are available (Sec. \ref{subsec:MLProblem_CG}).

In these examples, we have labeled training data $(\mathbf{x},\mathbf{f}(\mathbf{x}))_{i}$.
Using a network that directly predicts the forces would not guarantee
that Eq. (\ref{eq:conservative_force}) holds. Therefore, when Eq.
(\ref{eq:conservative_force}) holds, it is common to learn an energy
$U(\mathbf{x})$ and then computing the force in the network using
a gradient layer \citep{Nipsschnet,WangEtAl_ACSCS19_CGnet} (Fig.
\ref{fig:CGnet}a). An alternative to ensure Eq. (\ref{eq:conservative_force})
is gradient-domain machine learning \citep{ChmielaEtAl_SciAdv17_EnergyConserving}.

\subsubsection{Probability conservation and stochasticity}

\label{subsec:Probability-conservation-stochasticity}

In statistical mechanics (both equilibrium and non-equilibrium), we
are interested in the probability of events. An important principle
is thus probability conservation, i.e. that the sum over all events
is 1. A common approach to encode the probability of classes or events
with a neural network is to use a SoftMax output layer, where the
activation of each neuron can be defined as
\begin{equation}
y_{i}(\mathbf{u})=\frac{\mathrm{e}^{-u_{i}}}{\sum_{j}\mathrm{e}^{-u_{j}}}\label{eq:SoftMax}
\end{equation}
where $j$ runs over all neurons in the layer. In this representation,
$\mathbf{u}$ can be seen as a vector of energies giving rise to the
Boltzmann probabilities $y_{i}(\mathbf{u})$. $y_{i}(\mathbf{u})$
are nonnegative because of the exponential functions and sum up to
$1$.

In Markov state models (MSMs) of molecular kinetics, we would like
to obtain a Markov transition matrix $\mathbf{P}$ which is stochastic,
i.e. $p_{ij}\ge0$ for all elements and $\sum_{j}p_{ij}=1$ for all
$i$. If the encoder $E(\mathbf{x})$ uses a SoftMax, then the estimator
(\ref{eq:transition_matrix_estimator-1}) will result in a transition
matrix $\mathbf{P}$ that conserves probability ($\sum_{j}p_{ij}=1$
for all $i$). SoftMax can be exploited in VAMPnets to simultaneously
learn an embedding $E(\mathbf{x})$ from configurations to metastable
states, and a Markov transition matrix $\mathbf{P}$ (Sec. \ref{subsec:VAMPnets})
\citep{MardtEtAl_VAMPnets}.

\subsubsection{Detailed balance}

\label{subsec:Detailed-balance}

Detailed balance connects thermodynamics and dynamics. In a dynamical
system that evolves purely in thermal equilibrium, i.e. without applying
external forces, the equilibrium distribution $\mu(\mathbf{x})$ and
the transition probability $p(\mathbf{y}\mid\mathbf{x})$ are related
by the following symmetry:

\begin{equation}
\mu(\mathbf{x})p(\mathbf{y}\mid\mathbf{x})=\mu(\mathbf{y})p(\mathbf{x}\mid\mathbf{y}).\label{eq:detailed_balance}
\end{equation}
Thus, the unconditional probabilities of forward- and backward-trajectories
are equal. Enforcing Eq. (\ref{eq:detailed_balance}) in kinetic ML
models ensures the spectral decomposition (\ref{eq:density_propagation})
is real-valued, which is useful for many analyses.

In order to learn a kinetic model that obeys detailed balance, other
estimators than Eq. (\ref{eq:transition_matrix_estimator-1}) must
be used, that typically enforce the unconditional transition probabilities
$p(\mathbf{x},\mathbf{y})=\mu(\mathbf{x})p(\mathbf{y}\mid\mathbf{x})$
to be symmetric. See \citep{Noe_LectureNotes_ML4MD} for details.

\subsection{Parameter sharing and convolutions}

\label{subsec:Parameter-sharing}

A decisive advance in the performance of neural networks for classical
computer vision problems such as text or digit recognition came with
going from fully connected ``dense'' networks to convolutional neural
networks (CNN) \citep{LeCun_ProcIEEE89_ConvNets}. Convolution layers
are not only equivariant, and thus help with the detection of an object
independent of its location (Fig. \ref{fig:physical_constraints}c),
the real efficiency of CNNs is due to parameter sharing.

In a classical dense network, all neurons of neighboring layers are
connected and have independent parameters stored in a weight matrix
$\mathbf{W}^{l}$. This leads to a problem with high-dimensional data.
If we were to process relatively small images, say $100\times100=10^{4}$
pixels, and associate each pixel with a neuron, a single dense neural
network layer $l$ will have $10^{4}\times10^{4}=10^{8}$ parameters
in $\mathbf{W}^{l}$. This would not only be demanding in terms of
memory and computing time, a network with so many independent parameters
would likely overfit and not be able to generalize to unknown data.

Convolutions massively reduce the number of independent parameters.
A convolutional layer with a filter $\mathbf{w}$ acting on a one-dimensional
signal $\mathbf{x}^{l-1}$ computes, before applying bias and nonlinearities,
\begin{equation}
z_{i}^{l}=\sum_{j}x_{i-j}^{l-1}w_{j}^{l}.\label{eq:ConvLayer}
\end{equation}
In terms of an image, convolution applies the same filter $\mathbf{w}$
to every pixel neighborhood, in other words, all pixel transformations
share the same parameters. Additionally, the filters $\mathbf{w}$
are usually much smaller than the signal $\mathbf{x}$ (often $3\times3$
for images), thus each convolution only has few parameters. For molecules,
we extend this idea to continuous convolutions that use particle positions
instead of pixels (Sec. \ref{subsec:SchNet}).

Besides the sheer reduction of parameters, parameter sharing, e.g.
via convolution layers, is the keystone of transferability across
chemical space. In convolutions of a greyscale image, we apply the
same filters to every pixel, which implies translational equivariance,
but also means ``the same rules apply to all pixels''. If we have
multiple color channels, we have different channels in the filters
as well, so same color channels behave the same. In molecules we can
apply the same idea to particle species. When learning energies from
QM data, for example, every chemical element should be treated the
same, i.e. use the same convolution filters in order to sense its
chemical environment. This treatment gives us a ``building block''
principle that allows us to train on one set of molecules and make
predictions for new molecules.

\section{Deep Learning Architectures for Molecular Simulation}

\label{sec:Architectures}

In this section, we present specific methods and neural network architectures
that have been proposed to tackle the machine learning problems discussed
in Sec. \ref{sec:ML_problems}.

\subsection{Behler-Parrinello, ANI, Deep Potential net}

\label{subsec:DeepLearningMD}

Behler-Parrinello networks are one of the first applications of machine
learning in the molecular sciences \citep{BehlerParrinello_PRL07_NeuralNetwork}.
They aim at learning and predicting potential energy surfaces from
QM data and combine all of the relevant physical symmetries and parameter
sharing for this problem (Sec. \ref{sec:Physics}).

\begin{figure}
\begin{centering}
\includegraphics[width=0.8\columnwidth]{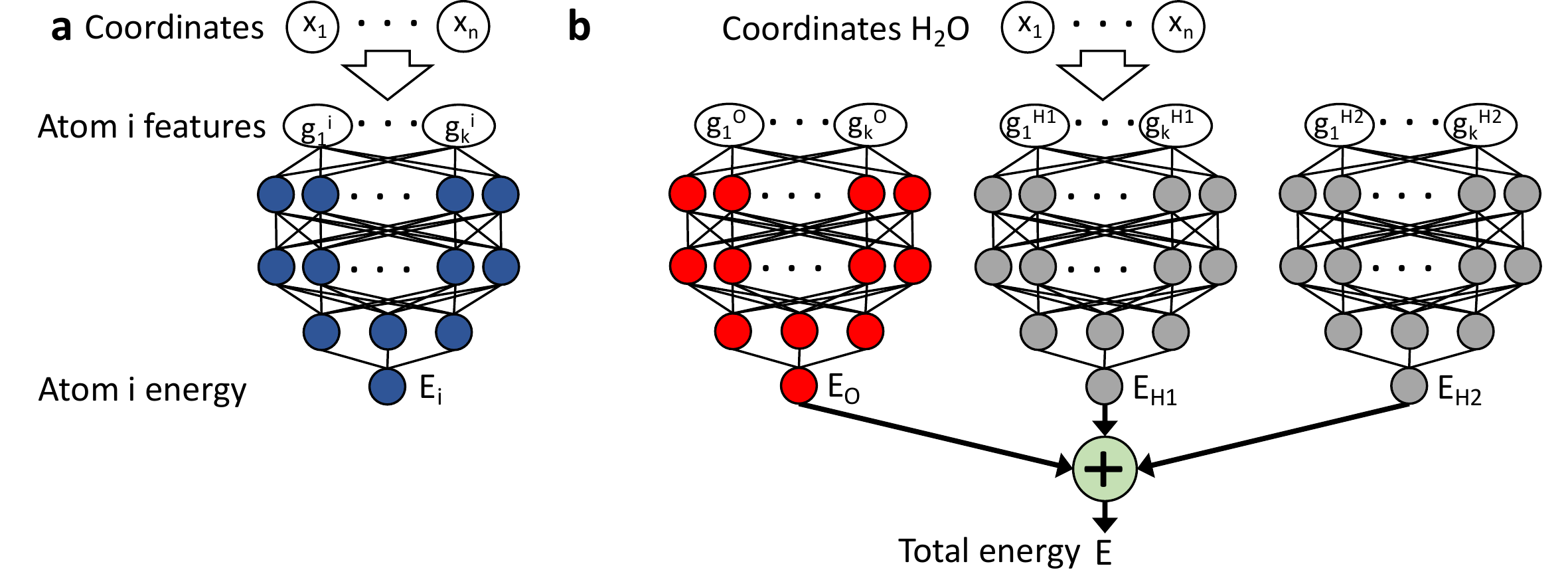}
\par\end{centering}
\caption{\label{fig:Behler-Parrinello}\textbf{Behler-Parrinello networks for
learning quantum energies}. \textbf{a}) Behler-Parrinello network
computing the atomic energy of a single atom $i$. The system coordinates
are mapped to roto-translationally invariant features describing the
chemical environment of atom $i$. \textbf{b}) A molecular system,
e.g., $H_{2}O$, is composed by employing a network copy for each
atom. The parameters are shared between networks for same elements.
The total energy is given by the sum of atomic energies, introducing
permutation invariance.}
\end{figure}

First, the molecular coordinates are mapped to roto-translationally
invariant features for each atom $i$ (Fig. \ref{fig:Behler-Parrinello}a).
In this step, the distances of neighboring atoms of a certain type,
as well as the angles between two neighbors of certain types are mapped
to a fixed set of correlation functions that describes the chemical
environment of atom $i$. These features are the input to a dense
neural network which outputs one number, the energy of atom $i$ in
its environment. By design of the input feature functions, this energy
is roto-translationally invariant. Parameters are shared between equivalent
atoms, e.g., all carbon atoms are using the same network parameters
to compute their atomic energies, but since the chemical environments
will differ, their energies will differ. In a second step, the atomic
energies are summed over all atoms of the molecule (Fig. \ref{fig:Behler-Parrinello}b).
This second step, combined with parameter sharing, achieves permutation
invariance, as explained in Sec. \ref{subsec:Permutational_invariance}.
Transferability is achieved due to parameter sharing, but also because
the summation principle allows to grow or shrink the network to molecules
of any size, including sizes that were never seen in the training
data.

A related approach is Deep-Potential net \citep{HanZhangCarE_PRL18_DeepPotNet},
where each atom is treated in a local coordinate frame that has the
rotation an translation degrees of freedom removed.

Behler-Parrinello networks are traditionally trained by energy matching
(Sec. \ref{subsec:MLProblem_PES}) \citep{BehlerParrinello_PRL07_NeuralNetwork,Behler-review-JCP},
but can be trained with force matching if a gradient layer is added
to compute the conservative force (Sec. \ref{subsec:MLProblem_PES},
Eq. \ref{eq:conservative_force}). The Behler-Parrinello method has
been further developed in the ANI network, e.g., by extending to more
advanced functions involving two neighbors \citep{SmithIsayevRoitberg_ChemSci17_ANI}.While
Behler-Parrinello networks have mainly been used to make predictions
of the same molecular system in order to run MD simulations unaffordable
by direct ab-initio QM MD \citep{Behler-review-JCP}, ANI has been
trained on DFT and coupled-cluster data across a large chemical space
\citep{SmithIsayevRoitberg_ChemSci17_ANI,SmithIsayevRoitberg_SciData17_ANIdata,SmithEtAl_ANI-CC}.

\subsection{Deep Tensor Neural Nets, SchNet and continuous convolutions}

\label{subsec:SchNet}

One of the first deep learning architectures to learn to represent
molecules or materials is the family of Deep Tensor Neural Networks
(DTNN) \citep{DTNN}, with its recent addition SchNet \citep{SchNet-JCP,Schnetsoftwarepaper}.
While in kernel-based learning methods \citep{Vapnik95,Muelleretal2001}
chemical compounds are compared in terms of pre-specified kernel functions
\citep{RuppEtAl_PRL12_QML,HansenJCTC13,HansenJPCL,BartokKondorCsanyi_PRB13_SOAP},
DTNN and its extension SchNet learn a multi-scale representation of
the properties of molecules or materials from large data sets.

\begin{figure}
\begin{centering}
\includegraphics[width=1\columnwidth]{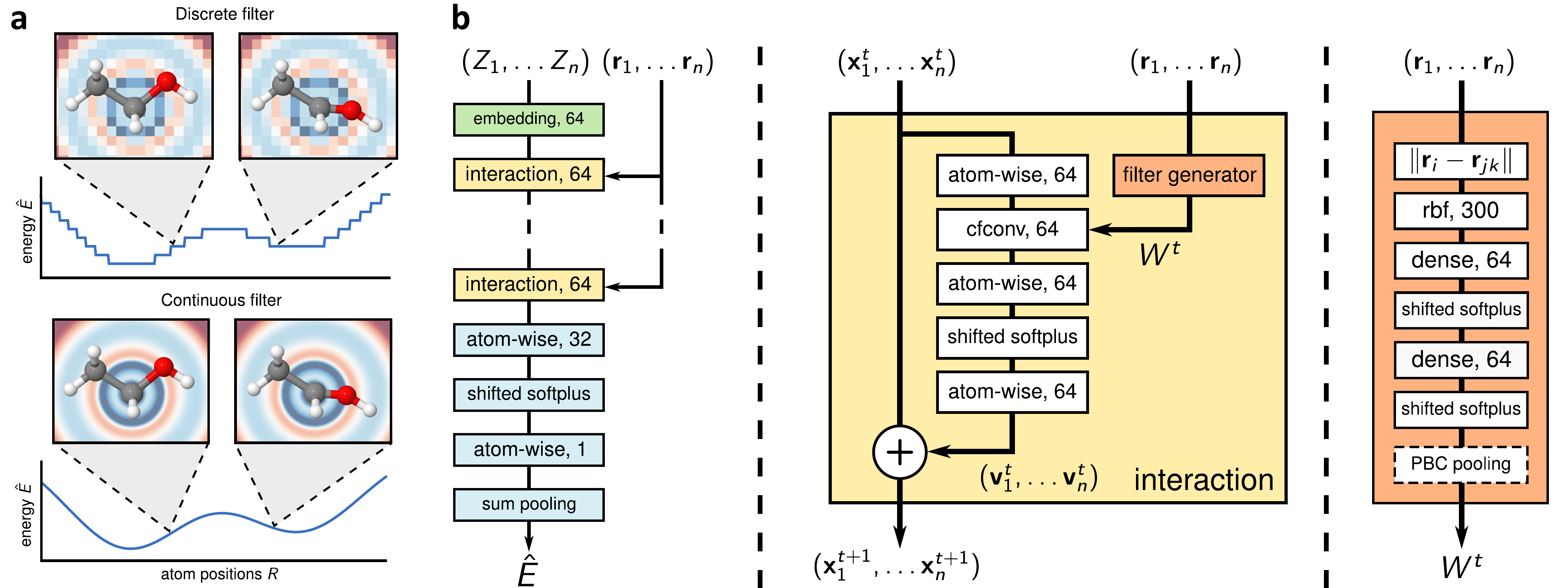}
\par\end{centering}
\caption{\textbf{\label{fig:SchNet}SchNet}, a continuous convolution framework:
\textbf{a}) As molecular structures cannot be well discretized on
a grids, SchNet generalizes the ConvNets approach to continuous convolutions
between particles. \textbf{b}) SchNet architecture: the input, consisting
of atom types (chemical elements $Z_{1},...,Z_{n}$) and positions
$\mathbf{r}_{1},...,\mathbf{r}_{n}$ is processed through several
layers to produce atom-wise energies that are summed to a total energy
(left). The most important layer is the interaction layer in which
atoms interact via continuous convolution functions (middle). Continuous
convolutions are generated by dense neural networks that operate on
the interatomic distances, ensuring roto-translational invariance
of the energy.}
\end{figure}

DTNNs were inspired by the language processing approach word-to-vec
\citep{mikolov2013distributed}, where the role of a word within its
grammatical or semantic context are learned and encoded in a parameter
vector. Likewise, DTNNs learn a representation vector for each atom
within its chemical environment (Fig. \ref{fig:SchNet}b left). DTNN\textquoteright s
tensor construction algorithm then iteratively learns higher-order
representations by first interacting with all pairwise neighbors,
e.g. extracting information implemented in the bond structure (Fig.
\ref{fig:SchNet}b middle). By stacking such interaction layers deep,
DTNNs can represent the structure and statistics of multi-body interactions
in deeper layers. As DTNNs are end-to-end trained to predict certain
quantum mechanical quantities, such as potential energies, they learn
the representation that is relevant for the task of predicting these
quantities \citep{Braun2008}.

SchNet \citep{SchNet-JCP} uses a deep convolutional neural network
(CNN)\citep{LeCun89,LeCun1998}. Classically, CNNs were developed
for computer vision using pixelated images, and hence use discrete
convolution filters. However, the particle positions of molecules
cannot be discretized on a grid as quantum mechanical properties such
as the energy are highly sensitive to small position changes, such
as the stretching of a covalent bond. For this reason, SchNet introduced
continuous convolutions \citep{SchNet-JCP,Nipsschnet}, which are
represented by filter-generating neural networks that map the roto-translationally
invariant interatomic distances to the filter values used in the convolution
(Fig. \ref{fig:SchNet}b right).

DTNN and SchNet have both reached highly competitive prediction quality
both across chemical compound space and across configuration space
in order to simulate molecular dynamics. In addition to their prediction
quality, their scalability to large data sets \citep{Schnetsoftwarepaper}
and their ability to extract novel chemical insights by means of their
learnt representation make the DTNN family an increasingly popular
research tool.

\subsection{Coarse-graining: CGnets}

\label{subsec:CGnets}

\begin{figure}[t]
\begin{centering}
\includegraphics[width=1\columnwidth]{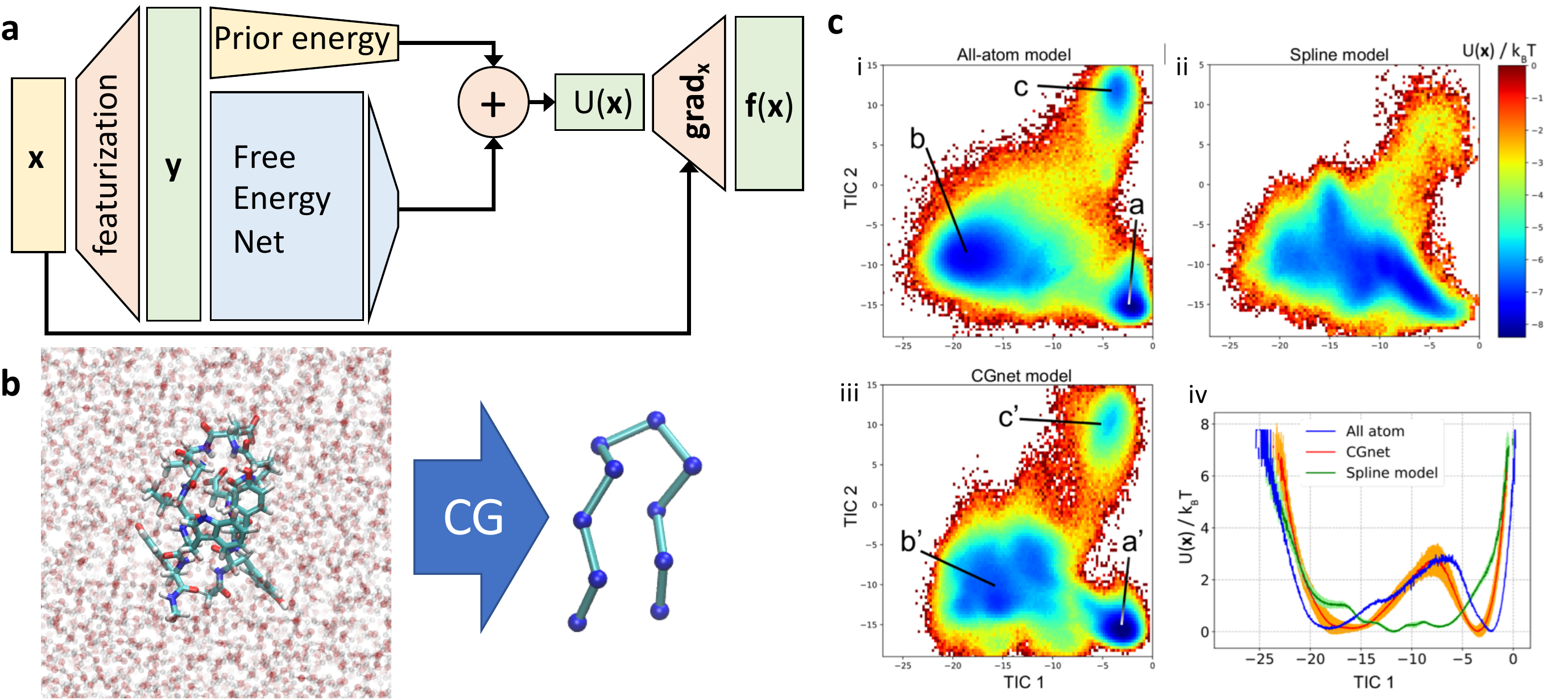}
\par\end{centering}
\caption{\label{fig:CGnet}\textbf{CGnet}. \textbf{a}) The CGnet neural network
architecture can be used to design a coarse-grained model by force-matching,
by using the loss function of Eq. (\ref{eq:loss-force-matching_linear}).
\textbf{b}) Application of CGnet to the coarse-graining of the small
protein Chignolin, from the fully atomistic and solved model to a
$C_{\alpha}$-only model of 10 beads. \textbf{c}) Results of CGnet
for Chignolin. i) The free energy of the original atomistic model,
as a function of the two reaction coordinates. State a is the folded
state, b is the unfolded state, and c is a misfolded state. ii) The
free energy resulting from a coarse-grained model where only two-body
terms are included in the energy function. State a', b', and c' correspond
to states a, b, and c in i). iii) The free energy resulting from CGnet.
iv) The same free energies as in i), ii), and iii) but as a function
of only one reaction coordinate. Figures adapted from \citep{WangEtAl_ACSCS19_CGnet}.}
\end{figure}
As mentioned in section \ref{subsec:MLProblem_CG}, machine learning
has been used to define coarse-grained models for molecular systems.
Both kernel methods \citep{John2017} and deep neural networks \citep{ZhangEtAl_JCP18_DeePCG,WangEtAl_ACSCS19_CGnet}
have been designed to minimize the force-matching loss, Eq. (\ref{eq:loss-force-matching_linear}),
for given coarse-graining mappings for specific systems.

In both cases, it has been shown that the incorporation of physical
constraints is crucial to the success of the model. The training data
are obtained by means of atomistic molecular dynamic simulations and
regions of the configurational space that are physically forbidden,
such as configurations with broken covalent bonds or overlapping atoms,
are not sampled and not included in the training. Without additional
constraints, the machine cannot make predictions far away from the
training data, and will thus not reliably predict that the energy
should diverge when approaching physically forbidden regions.

Excluding high-energy states such as broken bonds or colliding atoms
is different from enforcing physical symmetries as described in Sec.
\ref{subsec:Invariance-and-Equivariance}. Rather, it is about enforcing
the correct asymptotic behavior of the energy when going towards an
unphysical limit. CGnets proposed to achieve this by learning the
difference to a simple prior energy that was defined to have the correct
asymptotic behavior \citep{WangEtAl_ACSCS19_CGnet} (Fig. \ref{fig:CGnet}a).
The exact form of this prior energy is not essential for success,
as the CGnet can correct the prior energy where training data is available.
In \citep{WangEtAl_ACSCS19_CGnet}, the prior energy consisted of
harmonic terms for bonds and angles of coarse-grained particles whose
equilibrium values and force constants were obtained with Boltzmann
inversion, as well as excluded volume terms in the form of $\left(\sigma/r\right)^{c}$
were $r$ is the inter-particle distance and $\sigma,c$ are hyper-parameters.

As in Behler-Parrinello networks and SchNet, CGnet predicts a roto-translationally
invariant energy as the first layer transforms the Cartesian coordinates
into internal coordinates such as distances and angles (Fig. \ref{fig:CGnet}a).
Furthermore, CGnet predicts a conservative and rotation-equivariant
force field as the gradient of the total free energy with respect
to input configuration $\mathbf{x}$ is computed self-consistently
by the network (see Fig. \ref{fig:CGnet}a). The network is trained
by minimizing the force matching loss of this prediction (Eq. \ref{eq:loss_force_matching2}).

Fig. \ref{fig:CGnet}b-c shows an application of CGnets to the coarse-graining
of the mini-protein Chignolin, in which all solvent molecules are
coarse-grained away and the atoms of each residue are mapped to the
corresponding $C_{\alpha}$ atom (Fig. \ref{fig:CGnet}b). MD simulations
performed with the force field predicted by the CGnet predicts a free
energy surface that is quantitatively similar to the free energy surface
of the all-atom simulations, and resolves the same metastable states
(folded, unfolded and misfolded). In contrast, a spline-based coarse-grained
model where only two-body terms are included in the energy function
cannot reproduce the all-atom free energy surface, and does not even
predict that folded and unfolded are separated metastable states.
These results clearly illustrate the importance of multi-body interactions
in the coarse-grained energy, for example surface or volume terms
that can describe implicit solvation. While the spline model can be
dramatically improved by adding suitable terms to list of features,
this is not necessary when using a deep neural network which automatically
learns the required multi-body terms. Similar conclusions can be obtained
by using Gaussian Approximation Potentials as the machine learning
model to capture multi-body terms in coarse-grained energy functions
\citep{John2017}.

\subsection{Kinetics: VAMPnets}

\label{subsec:VAMPnets}

VAMPnets \citep{MardtEtAl_VAMPnets} were introduced to replace the
complicated and error-prone approach of constructing Markov state
models by (i) searching for optimal features, (ii) combining them
into a low-dimensional representation $\mathbf{y}$, e.g., via TICA
\citep{PerezEtAl_JCP13_TICA}, (iii) clustering $\mathbf{y}$, (iv)
estimating the transition matrix $\mathbf{P}$, and (v) coarse-graining
$\mathbf{P}$. VAMPnets uses instead a single end-to-end learning
approach in which all of these steps are replaced by a deep neural
network. This is possible because with the VAC and VAMP variational
principles (Sec. \ref{subsec:MLProblem_Kinetics}, \citep{NoeNueske_MMS13_VariationalApproach,WuNoe_VAMP}),
loss functions are available that are suitable to train the embedding
$E(\mathbf{x})$ and the propagator $\mathbf{P}$ simultaneously (see
Sec. \ref{subsec:MLProblem_Kinetics}, Eq. \ref{eq:learningDynamics}).

\begin{figure}[H]
\begin{centering}
\includegraphics[width=0.9\columnwidth]{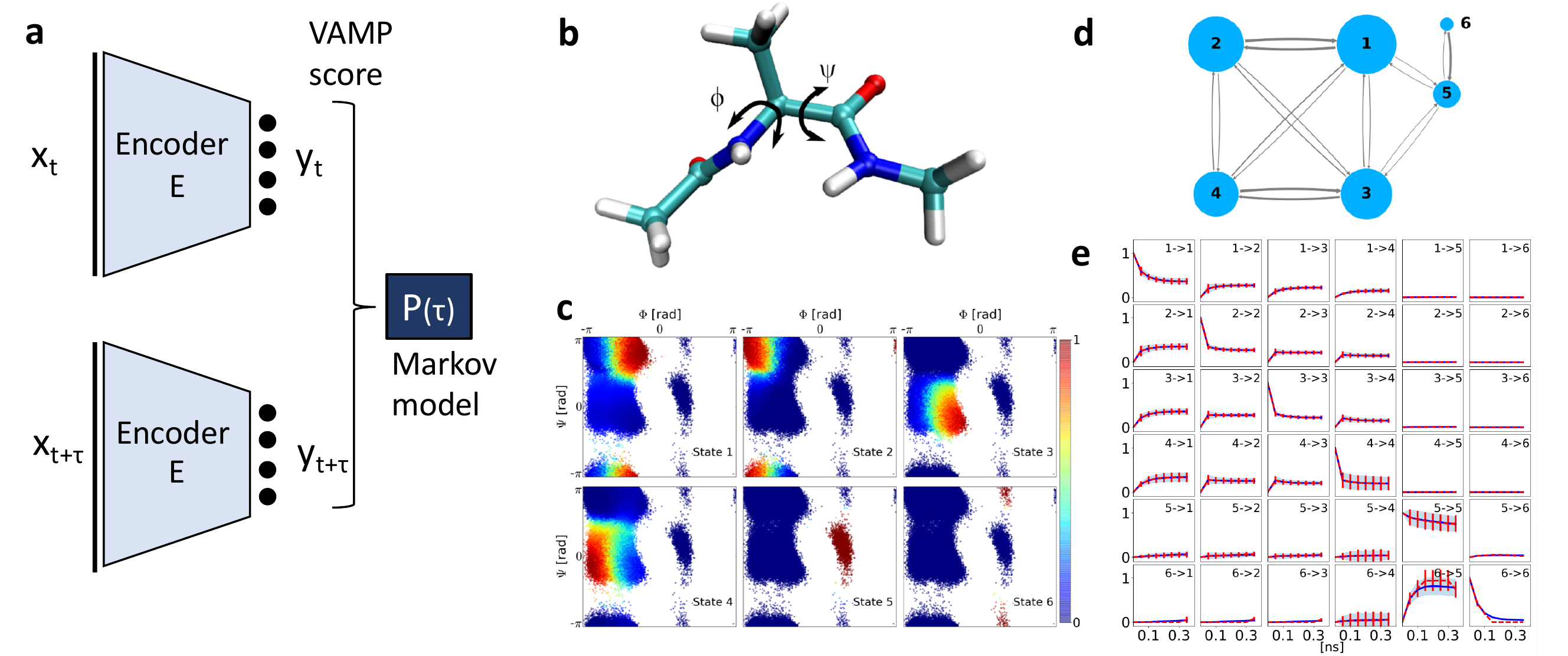}
\par\end{centering}
\caption{\label{fig:VAMPnet}\textbf{VAMPnet and application to alanine dipeptide}.
\textbf{a}) a VAMPnet \citep{MardtEtAl_VAMPnets} includes an encoder
$E$ which transforms each molecular configuration $\mathbf{x}_{t}$
to a latent space of ``slow reaction coordinates'' $\mathbf{y}_{t}$,
and is trained on pairs $(\mathbf{y}_{t},\mathbf{y}_{t+\tau})$ sampled
from the MD simulation using the VAMP score \citep{WuNoe_VAMP}. If
the encoder performs a classification, the dynamical propagator $\mathbf{P}(\tau)$
is a Markov state model. \textbf{b}) Structure of alanine dipeptide.
The backbone torsion angles $\phi$ and $\psi$, describe the slow
kinetics of conformation changes, but Cartesian coordinates of heavy
atoms are used as VAMPnet inputs here. \textbf{c}) Classification
of the VAMPnet encoder of MD frames to metastable states in the $\phi$
and $\psi$ space. Color corresponds to activation of the respective
output neuron. \textbf{d}) equilibrium probabilities of Markov states
and transition probabilities given by $\mathbf{P}(\tau)$. \textbf{e})
Chapman--Kolmogorov test comparing long-time predictions of the VAMPnet
model estimated at $\tau=50\:\mathrm{ps}$ with estimates at longer
lag times. Figure modified from \citep{MardtEtAl_VAMPnets}.}
\end{figure}

VAMPnets contain two network lobes representing the embedding $E(\mathbf{x})$.
These networks transform the molecular configurations found at a time
delay $\tau$ along the simulation trajectories (Fig. \ref{fig:VAMPnet}a).
VAMPnets can be trained by minimizing the VAMP loss, Eq. (\ref{eq:VAMP2_loss_b}),
which is meaningful for both dynamics with and without detailed balance
\citep{WuNoe_VAMP}. VAMPnets may, in general, use two distinct network
lobes to encode the spectral representation of the left and right
singular functions (which is important for non-stationary dynamics
\citep{KoltaiCiccottiSchuette_JCP16_MetastabilityNonstationary,KoltaiEtAl_Computation18_NonrevMSM}).
EDMD with dictionary learning \citep{LiEtAl_Chaos17_EDMD_DL} uses
a similar architecture as VAMPnets, but is optimized by minimizing
the regression error in latent space. In order to avoid collapsing
to trivial embeddings such as constant functions (see Sec. \ref{subsec:MLProblem_Kinetics})
a suitable regularization must be employed \citep{LiEtAl_Chaos17_EDMD_DL}.

While hyper-parameter selection can be performed by minimizing the
variational loss (\ref{eq:VAMP2_loss_b}) on a validation set \citep{BiessmannEtAl_ML10_TCCA,McGibbonPande_JCP15_CrossValidation,HusicEtAl_JCP16_GMRQTrends,SchererEtAl_VAMPselection},
it is important to test the performance of a kinetic model on timescales
beyond the training timescale $\tau$. We can use the Chapman-Kolmogorov
equation to test how well the learnt model predicts longer times:
\begin{equation}
\mathbf{P}^{n}(\tau)\approx\mathbf{P}(n\tau).\label{eq:CKtest}
\end{equation}
A common way to implement this test is to compare the leading eigenvalues
$\lambda_{i}(\tau)$ of the left and right hand sides \citep{SwopePiteraSuits_JPCB108_6571,NoeSchuetteReichWeikl_PNAS09_TPT,PrinzEtAl_JCP10_MSM1}.

In \citep{MardtEtAl_VAMPnets}, parameters were shared between the
VAMPnets nodes, and thus a unique embedding $E(\mathbf{x})$ is learned.
When detailed balance is enforced while computing $\mathbf{P}(\tau)$
(Eq. \ref{eq:detailed_balance}), the loss function automatically
becomes a VAC score \citep{NoeNueske_MMS13_VariationalApproach}.
In this case, the embedding $E(\mathbf{x})$ encodes the space of
the dominant Markov operator eigenfunctions \citep{MardtEtAl_VAMPnets}.
This feature was extensively studied in state-free reversible VAMPnets
\citep{CheeSidkyFerguson_arxiv19_ReversibleVAMPnets}.

In order to obtain a propagator that can be interpreted as a Markov
state model, \citep{MardtEtAl_VAMPnets} chose to use a SoftMax layer
as an output layer, thus transforming the spectral representation
to a soft indicator function similar to spectral clustering methods
such as PCCA+ \citep{DeuflhardWeber_LAA05_PCCA+,RoeblitzWeber_AdvDataAnalClassif13_PCCA++}.
As a result, the propagator computed by Eq. (\ref{eq:transition_matrix_estimator-1})
conserves probability and is almost a transition matrix (Sec. \ref{subsec:Probability-conservation-stochasticity}),
although it may have some negative elements with small absolute values.

The results described in \citep{MardtEtAl_VAMPnets} (see, e.g., Fig.
\ref{fig:VAMPnet}) were competitive with and sometimes surpassed
the state-of-the-art handcrafted MSM analysis pipeline. Given the
rapid improvements of training efficiency and accuracy of deep neural
networks seen in a broad range of disciplines, we expect end-to-end
learning approaches such as VAMPnets to dominate the field eventually.

In \citep{WuMardtPasqualiNoe_DeepGenerativeMSMs}, a deep generative
Markov State Model (DeepGenMSM) was proposed that, in addition to
the encoder $E(\mathbf{x})$ and the propagator $\mathbf{P}$ learns
a generative part that samples the conditional distribution of configurations
in the next time step. The model can be operated in a recursive fashion
to generate trajectories to predict the system evolution from a defined
starting state and propose new configurations. The DeepGenMSM was
demonstrated to provide accurate estimates of the long-time kinetics
and generate valid distributions for small molecular dynamics benchmark
systems.

\subsection{Sampling/Thermodynamics: Boltzmann Generators}

\label{subsec:BGs}

Boltzmann Generators were introduced in \citep{NoeEtAl_19_BoltzmannGenerators}
to learn to sample equilibrium distributions, Eq. (\ref{eq:equilibrium}).
In contrast to standard generative learning, a Boltzmann Generator
does not attempt to learn the probability density from data, but is
trained to efficiently sample $\mu(\mathbf{x})\propto\mathrm{e}^{-u(\mathbf{x})}$
using the dimensionless energy $u(\mathbf{x})$ as an input. A Boltzmann
Generator consists of two parts:
\begin{enumerate}
\item A generative model that is trained to propose samples from a probability
distribution $p_{X}(\mathbf{x})$ which is ``similar'' to $\mu(\mathbf{x})$,
and that allows us to evaluate $p_{X}(\mathbf{x})$ (up to a constant)
for every $\mathbf{x}$.
\item A reweighting procedure that takes proposals from $p_{X}(\mathbf{x})$
and generates unbiased samples from $\text{\ensuremath{\mu}(\ensuremath{\mathbf{x}})}$.
\end{enumerate}
Boltzmann Generators use a trainable generative network $F_{zx}$
which maps latent space samples $\mathbf{z}$ from a simple prior,
e.g., a Gaussian normal distribution, to samples $\mathbf{x}\sim p_{X}(\mathbf{x})$.
Training is done by combining the energy-based training using the
KL divergence, Eq. (\ref{eq:equilibrium_KLloss}), and maximum likelihood,
Eq. (\ref{eq:equilibrium_MLloss}).

\begin{figure}[t]
\begin{centering}
\includegraphics[width=0.8\columnwidth]{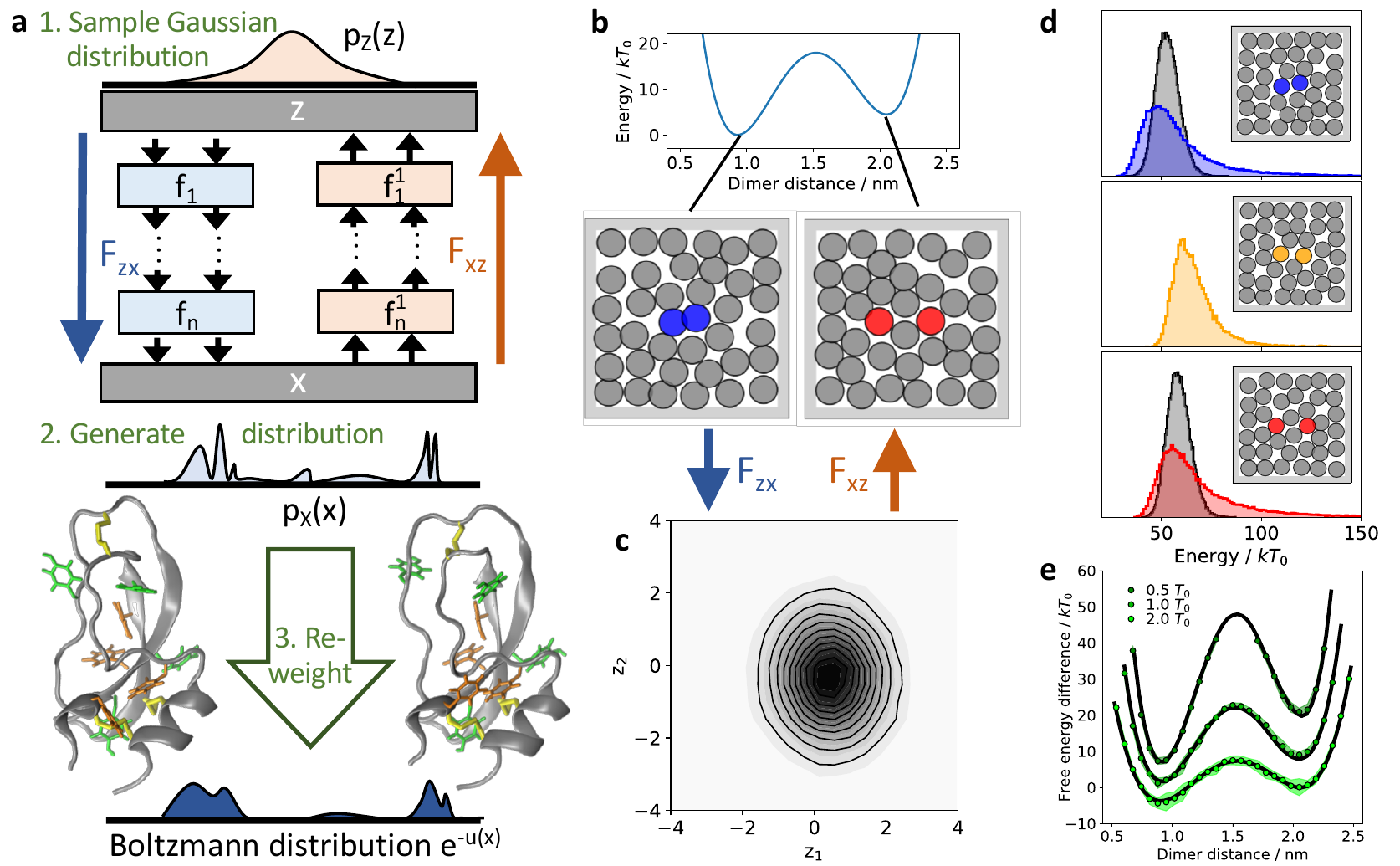}
\par\end{centering}
\caption{\label{fig:BGs}\textbf{Boltzmann Generators}: \textbf{a}) A Boltzmann
Generator is trained by minimizing the difference between its generated
distribution and the desired Boltzmann distribution. Generation proceeds
by drawing ``latent'' space samples $\mathbf{z}$ from a simple
prior distribution (e.g., Gaussian) and transforming them to configurations
$\mathbf{x}$ via an invertible Flow: a deep neural network $F_{zx}$
and its inverse, $F_{xz}$. To compute thermodynamics, such as configurational
free energies, the samples must be reweighted to the Boltzmann distribution.
\textbf{b}) Repulsive particle system with bistable dimer. Closed
(blue) and open (red) configurations from MD simulations (input data).
\textbf{c}) Distribution of MD simulation data in latent space coordinates
$z_{1},z_{2}$ after training Boltzmann Generator. \textbf{d}) Potential
energy distribution from MD (grey) and Boltzmann generator for closed
(blue), open (red) and transition configurations (yellow). Insets
show one-shot Boltzmann Generator samples. \textbf{e}) Free energy
differences as a function of dimer distance and relative temperature
sampled with Boltzmann generators (green bullets) and umbrella sampling
(black lines). Figure modified from \citep{NoeEtAl_19_BoltzmannGenerators}.}
\end{figure}

For both, training and reweighting, we need to be able to compute
the probability $p_{X}(\mathbf{x})$ of generating a configuration
$\mathbf{x}$. This can be achieved by the change-of-variables equation
if $F_{zx}$ is an invertible transformation (Fig. \ref{fig:BGs}a)
\citep{TabakVandenEijnden_CMS10_DensityEstimation,DinhDruegerBengio_NICE2015}.
Such invertible networks are called flows due to the analogy of the
transformed probability density with a fluid \citep{DinhDruegerBengio_NICE2015,DinhBengio_RealNVP}.
In \citep{NoeEtAl_19_BoltzmannGenerators}, the non-volume preserving
transformations RealNVP were employed \citep{DinhBengio_RealNVP},
but the development of more powerful invertible network architectures
is an active field of research \citep{RezendeEtAl_NormalizingFlows,KingmaDhariwal_NIPS18_Glow,GrathwohlEtAl_FFJORD}.
By stacking multiple invertible ``blocks'', a deep invertible neural
network is obtained that can encode a complex transformation of variables.

Fig. \ref{fig:BGs}c-h illustrate the Boltzmann Generator on a condensed-matter
model system which contains a bistable dimer in a box densely filled
with repulsive solvent particles (Fig. \ref{fig:BGs}c,d). Opening
or closing the dimer is a rare event, and also involves the collective
rearrangement of the solvent particles due to the high particle density.
Using short MD simulation in the open and closed states as an initialization,
the Boltzmann Generator can be trained to sample open, closed and
the previously unseen transition states by generating Gaussian random
variables in latent space (Fig. \ref{fig:BGs}e) and feeding them
through the transformation $F_{zx}$. Such samples have realistic
structures and close to equilibrium energies (Fig. \ref{fig:BGs}f).
By employing reweighting, free energy differences can be computed
(Fig. \ref{fig:BGs}g). There is a direct relationship between the
temperature of the canonical ensemble and the variance of the latent-space
Gaussian of the Boltzmann Generator \citep{NoeEtAl_19_BoltzmannGenerators}.
This allows us to learn to generate thermodynamics, such as the temperature-dependent
free energy profiles, using a single Boltzmann Generator (Fig. \ref{fig:BGs}g).
Finally, as the latent space concentrates configurations of equilibrium
probability around the origin, Boltzmann Generators can be used to
generate physically realistic reaction pathways by performing linear
interpolations in latent space.

\section{Discussion}

\label{sec:Discussion}

Despite rapid advances in the field of Machine Learning for Molecular
Simulation, there are still significant open problems that need to
be addressed, in all the areas discussed above.

\subsection{Accuracy and efficiency in quantum-chemical energies and forces}

In order to be practically useful, a ML model for both PES and atomic
forces is needed that: i) can yield accuracy of $0.2-0.3$ kcal/mol
for the energy per functional group and about $1$ kcal/mol/Å for
the force per atom; ii) is not much more expensive to evaluate than
classical force fields, iii) scales to large molecules such as proteins;
and iv) is transferable to different covalent and non-covalent environments.
Such universal model does not exist yet.

Crucial steps towards i-ii) have been recently taken by symmetrized
gradient-domain machine learning (sGDML), a kernel-based approach
to constructing molecular force fields \citep{ChmielaEtAl_SciAdv17_EnergyConserving,ChmielaEtAl_NatComm18_TowardExact,sGDML-JCP,sGDML-CPC}.
Currently, sGDML already enables MD simulations with electrons and
nuclei treated at essentially exact quantum-mechanical level for molecules
with up to 20-30 atoms.

Network-based approaches, such as Schnet, ANI etc, are better suited
to iii-iv), as they break down the energy in local interactions of
atoms with their environment, thus enabling a ``building block''
principle that is by design better scalable to molecules of different
size and transferable across chemical space. However, these approaches
do not reach the high accuracy in configuration space that sGDML does.
Combining high accuracy in configuration and chemical space remains
an active research topic.

\subsection{Long-ranged interactions}

The vast majority of approaches to make ML inference on molecular
structures are based on local chemical information. Current neural
networks for modeling molecular energies use the summation principle
(e.g., Eq. \ref{eq:energy_atomwise}) in order to sum up local energies
$E_{i}(\mathbf{x})$ of atom $i$ with its neighbors. While multi-body
and long-ranged energies can be obtained by stacking multiple layers
\citep{DTNN,SchNet-JCP,Nipsschnet} -- the working principle of deep
convolution networks \citep{LeCun89} -- there are fundamental physical
limits of this approach: Long-ranged interactions such as electrostatics
cannot be cut off.

For classical point-charge models, long-ranged electrostatics methods
have been developed, such as the Ewald summation method for periodic
systems \citep{Darden_Structure99}. One option is to combine short-ranged
ML models with such long-ranged electrostatics methods. In order to
avoid double counting interactions, one must also predict atomic charges,
which is an active field of research \citep{UnkeMeuwli_PhysNet,NebgenEtAl_JCTC18_TransferableChargeLearning}.
An alternative option, and currently unexplored territory, is to develop
neural network structures for particle interactions that can compute
long-ranged interactions by design.

In addition to electrostatics, van der Waals (vdW) dispersion interactions
can also have a substantial long-range character, i.e. they can extend
to separations of tens of nanometers or more in large molecular and
nanoscale systems \citep{AmbrosettiTkatchenko_Science16_VdW,HermannEtAl_ChemRev17_VdWReview,StoehrEtAl_CSR19_VdWReview}.
Developing ML models that correctly treat the quantum-mechanical many-body
nature of vdW interactions remains a difficult challenge to overcome
\citep{BereauEtAl_JCP18_NoncovalentML}.

\subsection{Quantum Kinetics}

With the availability of chemically transferable ML models that have
quantum-chemical accuracy, the next open problem is to sample metastable
states and long timescale kinetics. Although available ML models for
predicting QM energies and forces are still significantly slower than
force fields, the vast array of enhanced sampling methods and kinetic
models (Sec. \ref{subsec:MLProblem_Kinetics},\ref{subsec:VAMPnets})
will likely allow us to explore kinetics of quantum chemical systems
on timescales of microseconds and beyond. A plethora of new physical
insights that we cannot access with current MD force fields awaits
us there. For example, what is the role of protonation dynamics in
mediating protein folding or function?

\subsection{Transferability of coarse-grained models}

An outstanding question in the design of coarse-grained models is
that of transferability across chemical space. Bottom-up coarse-grained
models are useful in practice if they can be parametrized on small
molecules and then used to predict the dynamics of systems much larger
than what is possible to simulate with atomistic resolution. It is
not clear to what extent transferability of coarse-grained models
can be achieved, and how that depends on the coarse-graining mapping
\citep{Mullinax2009b,Thorpe2011}. Compared to the manual design of
few-body free energy functionals, machine-learned free energies can
help with transferability, as they are able to learn the important
multi-body effects, e.g., to model neglected solvent molecules implicitly
(Sec. \ref{subsec:CGnets}) \citep{John2017,ZhangEtAl_JCP18_DeePCG,WangEtAl_ACSCS19_CGnet}.

It is natural to consider Behler-Parrinello type networks or SchNet
as a starting point for modeling transferable coarse-grained energies,
but their application is nontrivial: it is a priori unclear what the
interacting particles are in the coarse-grained model and how to define
their ``types'', as they are no longer given by the chemical element.
Furthermore, these networks assume permutation invariance between
identical particles, while classical MD force fields do not have permutation
invariance of atoms within the same molecule. Therefore, particle
network structures that can handle bonding graphs need to be developed.

\subsection{Kinetics of coarse-grained models}

While coarse-grained MD models may perform well in reproducing the
thermodynamics of the atomistic system, they may fail in reproducing
the kinetics. Existing approaches include adding fictitious particles
\citep{DyvtyanVothAnderson_JCP16_DynamicForceMatching}, or training
the coarse-grained model with spectral matching \citep{Nueske2019_SpectralMatching}.
There is indication that the kinetics can be approximately up to a
global scaling factor in barrier-crossing problems when the barriers
are well approximated \citep{BereauRudzinski_PRL18_CGKinetics}, which
could be achieved by identifying the slow reaction coordinates \citep{NoeClementi_COSB17_SlowCVs},
and assigning more weight to the transition state in force matching
or relative entropy minimization. This area of research is still underdeveloped.

\subsection{Transferable prediction of intensive properties}

Extensive properties such as potential energies can be well predicted
across chemical space, as they can be conceptually broken down as
a sum of parts that can be learnt separately. This is not possible
with intensive properties such as spectra or kinetics, and for this
reason the prediction of such properties is, as yet, far behind.

\subsection{Equivariant generative networks with parameter sharing}

Generative networks, such as Boltzmann Generators (Sec. \ref{subsec:BGs})
have been demonstrated to be able to generate physically realistic
one-shot samples of model systems and proteins in implicit solvent
\citep{NoeEtAl_19_BoltzmannGenerators}. In order to scale to larger
systems, important steps are to build the invariances of the energy,
such as the exchange of identical solvent particles, into the transformation,
and to include parameter sharing (Sec. \ref{subsec:Parameter-sharing}),
such that we can go beyond just sampling the probability density of
one given system with energy $u(\mathbf{x})$ and instead generalize
from a dataset of examples of one class of molecules, e.g. solvated
proteins. To this end, equivariant networks with parameter sharing
need to be developed for generative learning, which are, to date,
not available.

\subsection{Explainable AI}

Recently, the increasing popularity of explainable AI methods (see
e.g. \citep{Bach15,montavon2018methods,lapuschkin2019unmasking,SameketalBook19})
have allowed us to gain insight into the inner workings of deep learning
algorithms. In this manner, it has become possible to extract how
a problem is solved by the deep model. This allows for example to
detect so-called \textquoteleft clever Hans\textquoteright{} solutions
\citep{lapuschkin2019unmasking}, i.e. nonsensical solutions relying
on artifactual or nonphysical aspects in data. Combined with networks
that learn a representation such as DTNN/Schnet \citep{DTNN,Nipsschnet}
and VAMPnets \citep{MardtEtAl_VAMPnets}, these inspection methods
may provide scientific insights into the mechanisms that give rise
to the predicted physicochemical quantity, and thus fuel the development
of new theories.

\section*{Acknowledgements}

We gratefully acknowledge funding from European Commission (ERC CoG
772230 \textquotedblleft ScaleCell\textquotedblright{} to F.N. and
ERC-CoG grant BeStMo to A.T.), Deutsche Forschungsgemeinschaft (CRC1114/A04
to F.N., EXC 2046/1, Project ID 390685689 to K.-R.M., GRK2433 DAEDALUS
to F.N. and K.-R.M.), the MATH+ Berlin Mathematics research center
(AA1-8 to F.H., EF1-2 to F.N. and K.-R.M.), Einstein Foundation Berlin
(Einstein Visiting Fellowship to C.C.), the National Science Foundation
(grants CHE-1265929, CHE-1740990, CHE-1900374, and PHY-1427654 to
C.C.), the Welch Foundation (grant C-1570 to C.C.), the Institute
for Information \& Communications Technology Planning \& Evaluation
(IITP) grant funded by the Korea government (No. 2017-0-00451, No.
2017-0-01779 to K.-R.M.), and the German Ministry for Education and
Research (BMBF) (Grants 01IS14013A-E, 01GQ1115 and 01GQ0850 to K.-R.
M.). All authors thank Stefan Chmiela and Kristof Schütt for help
with Figures 1 and 5.

\bibliographystyle{unsrt}
\bibliography{all,own}

\end{document}